\documentclass[12pt,a4paper]{article}

\long\def\symbolfootnote[#1]#2{\begingroup%
\def\thefootnote{\fnsymbol{footnote}}\footnote[#1]{#2}\endgroup}

\usepackage{lineno}  
\usepackage{graphicx}  
\usepackage{xspace} 
\usepackage{color}
\usepackage{colortbl}
\usepackage{amsmath} 
\usepackage{ifthen} 
\usepackage{multirow}
\usepackage{dcolumn}
\usepackage{booktabs}
\usepackage{cite}

\newboolean{pdflatex}
\setboolean{pdflatex}{true} 
\newboolean{articletitles}
\setboolean{articletitles}{true} 

\newboolean{uprightparticles}
\setboolean{uprightparticles}{false} 

\usepackage{amssymb}
\usepackage{amsfonts}
\usepackage{upgreek} 

\usepackage{hyperref}    
\usepackage[all]{hypcap} 

\def\lhcb {\mbox{LHCb}\xspace}
\def\ux85 {\mbox{UX85}\xspace}

\def\lhc    {\mbox{LHC}\xspace}

\def\ecal   {ECAL\xspace}

\ifthenelse{\boolean{uprightparticles}}%
{
 
 \def\Pgamma      {\ensuremath{\upgamma}\xspace}

 \def\Ppi         {\ensuremath{\uppi}\xspace}

 \def\PDelta      {\ensuremath{\Delta}\xspace}                 
 \def\PXi      {\ensuremath{\Xi}\xspace}                 
 \def\PLambda      {\ensuremath{\Lambda}\xspace}                 
 \def\PSigma      {\ensuremath{\Sigma}\xspace}                 
 \def\POmega      {\ensuremath{\Omega}\xspace}                 
 \def\PUpsilon      {\ensuremath{\Upsilon}\xspace}                 
 
 \def\PB      {\ensuremath{\mathrm{B}}\xspace}                 
                  
 \def\PD      {\ensuremath{\mathrm{D}}\xspace}

 \def\PK      {\ensuremath{\mathrm{K}}\xspace}

 \def\Pe      {\ensuremath{\mathrm{e}}\xspace}

 \def\Pi      {\ensuremath{\mathrm{i}}\xspace}

 \def\Pn      {\ensuremath{\mathrm{n}}\xspace}                 
                  
 \def\Pp      {\ensuremath{\mathrm{p}}\xspace}

}
{
 
 \def\Pgamma      {\ensuremath{\gamma}\xspace}

 \def\Ppi         {\ensuremath{\pi}\xspace}

 \mathchardef\PDelta="7101
 \mathchardef\PXi="7104
 \mathchardef\PLambda="7103
 \mathchardef\PSigma="7106
 \mathchardef\POmega="710A
 \mathchardef\PUpsilon="7107
                  
 \def\PB      {\ensuremath{B}\xspace}                 
                  
 \def\PD      {\ensuremath{D}\xspace}

 \def\PK      {\ensuremath{K}\xspace}

 \def\Pe      {\ensuremath{e}\xspace}

 \def\Pi      {\ensuremath{i}\xspace}

 \def\Pn      {\ensuremath{n}\xspace}                 
                  
 \def\Pp      {\ensuremath{p}\xspace}

}

\def\epm        {\ensuremath{\Pe^\pm}\xspace}

\def\g      {\ensuremath{\Pgamma}\xspace}

\def\pion  {\ensuremath{\Ppi}\xspace}

\def\pipm  {\ensuremath{\pion^\pm}\xspace}

\def\kaon  {\ensuremath{\PK}\xspace}
  \def\Kbar  {\kern 0.2em\overline{\kern -0.2em \PK}{}\xspace}

\def\Kz    {\ensuremath{\kaon^0}\xspace}
\def\Kzb   {\ensuremath{\Kbar^0}\xspace}
\def\KzKzb {\ensuremath{\Kz \kern -0.16em \Kzb}\xspace}
\def\Kp    {\ensuremath{\kaon^+}\xspace}
\def\Km    {\ensuremath{\kaon^-}\xspace}
\def\Kpm   {\ensuremath{\kaon^\pm}\xspace}

\def\KpKm  {\ensuremath{\Kp \kern -0.16em \Km}\xspace}
 
\def\KL    {\ensuremath{\kaon^0_{\rm\scriptscriptstyle L}}\xspace}

  \def\Dbar    {\kern 0.2em\overline{\kern -0.2em \PD}{}\xspace}
\def\D       {\ensuremath{\PD}\xspace}

\def\Dz      {\ensuremath{\D^0}\xspace}
\def\Dzb     {\ensuremath{\Dbar^0}\xspace}
\def\DzDzb   {\ensuremath{\Dz {\kern -0.16em \Dzb}}\xspace}
\def\Dp      {\ensuremath{\D^+}\xspace}
\def\Dm      {\ensuremath{\D^-}\xspace}

\def\DpDm    {\ensuremath{\Dp {\kern -0.16em \Dm}}\xspace}

  \def\Bbar    {\kern 0.18em\overline{\kern -0.18em \PB}{}\xspace}

  \def\Y#1S{\ensuremath{\PUpsilon{(#1S)}}\xspace}

\def\proton      {\ensuremath{\Pp}\xspace}
\def\antiproton  {\ensuremath{\overline \proton}\xspace}
\def\neutron     {\ensuremath{\Pn}\xspace}
\def\antineutron {\ensuremath{\overline \neutron}\xspace}

\def\Lbar {\ensuremath{\kern 0.1em\overline{\kern -0.1em\PLambda}}\xspace}

\def\to                 {\ensuremath{\rightarrow}\xspace}

\def\AT#1     {\ensuremath{A_{\mathrm{T}}^{#1}}\xspace}           

\def\C#1      {\ensuremath{\mathcal{C}_{#1}}\xspace}                       
\def\Cp#1     {\ensuremath{\mathcal{C}_{#1}^{'}}\xspace}                    
\def\Ceff#1   {\ensuremath{\mathcal{C}_{#1}^{\mathrm{(eff)}}}\xspace}        
\def\Cpeff#1  {\ensuremath{\mathcal{C}_{#1}^{'\mathrm{(eff)}}}\xspace}       
\def\Ope#1    {\ensuremath{\mathcal{O}_{#1}}\xspace}                       
\def\Opep#1   {\ensuremath{\mathcal{O}_{#1}^{'}}\xspace}                    

\newcommand{\tev}{\ensuremath{\mathrm{\,Te\kern -0.1em V}}\xspace}
\newcommand{\gev}{\ensuremath{\mathrm{\,Ge\kern -0.1em V}}\xspace}
\newcommand{\mev}{\ensuremath{\mathrm{\,Me\kern -0.1em V}}\xspace}
\newcommand{\kev}{\ensuremath{\mathrm{\,ke\kern -0.1em V}}\xspace}
\newcommand{\ev}{\ensuremath{\mathrm{\,e\kern -0.1em V}}\xspace}
\newcommand{\gevc}{\ensuremath{{\mathrm{\,Ge\kern -0.1em V\!/}c}}\xspace}
\newcommand{\mevc}{\ensuremath{{\mathrm{\,Me\kern -0.1em V\!/}c}}\xspace}
\newcommand{\gevcc}{\ensuremath{{\mathrm{\,Ge\kern -0.1em V\!/}c^2}}\xspace}
\newcommand{\gevgevcccc}{\ensuremath{{\mathrm{\,Ge\kern -0.1em V^2\!/}c^4}}\xspace}
\newcommand{\mevcc}{\ensuremath{{\mathrm{\,Me\kern -0.1em V\!/}c^2}}\xspace}

\def\mum  {\ensuremath{\,\upmu\rm m}\xspace}

\def\invnb {\ensuremath{\mbox{\,nb}^{-1}}\xspace}

\def\gsim{{~\raise.15em\hbox{$>$}\kern-.85em
          \lower.35em\hbox{$\sim$}~}\xspace}
\def\lsim{{~\raise.15em\hbox{$<$}\kern-.85em
          \lower.35em\hbox{$\sim$}~}\xspace}

\def\sqs   {\ensuremath{\protect\sqrt{s}}\xspace}

\def\pt         {\mbox{$p_{\rm T}$}\xspace}

\def\mrad{\ensuremath{\rm \,mrad}\xspace}

\def\evtgen     {\mbox{\textsc{EvtGen}}\xspace}

\def\geant      {\mbox{\textsc{Geant4}}\xspace}

\def\photos     {\mbox{\textsc{Photos}}\xspace}

\def\tell1  {TELL1\xspace}
\def\ukl1   {UKL1\xspace}

\usepackage{mciteplus}

\begin{document}



\begin{titlepage}
\pagenumbering{roman}

\vspace*{-1.5cm}
\centerline{\large EUROPEAN ORGANIZATION FOR NUCLEAR RESEARCH (CERN)}
\vspace*{1.5cm}
\hspace*{-0.5cm}
\begin{tabular*}{\linewidth}{lc@{\extracolsep{\fill}}r}
\ifthenelse{\boolean{pdflatex}}
{\vspace*{-2.7cm}\mbox{\!\!\!\includegraphics[width=.14\textwidth]{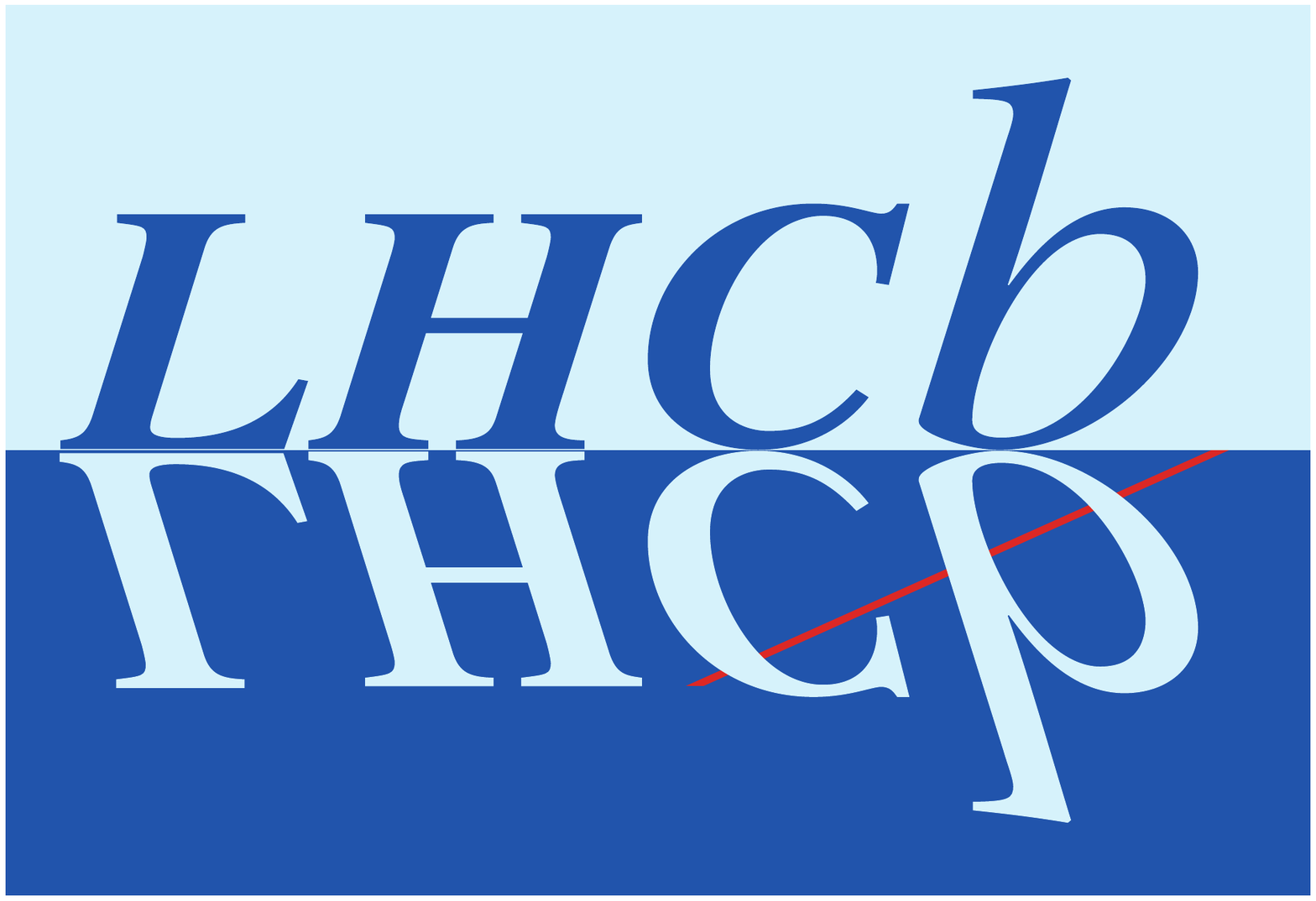}} & &}%
{\vspace*{-1.2cm}\mbox{\!\!\!\includegraphics[width=.12\textwidth]{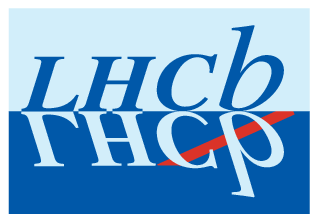}} & &}%
\\
 & & CERN-PH-EP-2012-346 \\  
 & & LHCb-PAPER-2012-034 \\  
 & & March 22, 2013              \\  
 & & \\
\end{tabular*}

\vspace*{4.0cm}

{\bf\boldmath\huge
\begin{center}
Measurement of the forward energy flow in $pp$ collisions 
at $\sqrt{s}=7$~TeV
\end{center}
}

\vspace*{2.0cm}

\begin{center}
The LHCb collaboration\footnote{Authors are listed on the following pages.}
\end{center}

\vspace{\fill}

\begin{abstract}
  \noindent
 The energy flow created in $pp$ collisions at $\sqrt{s}=7$~TeV is studied 
 within the pseudorapidity range $1.9<\eta<4.9$ with data collected by the \lhcb experiment. 
 The measurements are performed for inclusive minimum-bias interactions, hard scattering processes 
 and events with an enhanced or suppressed diffractive contribution. 
 The results are compared to predictions given by {\sc Pythia}-based and 
 cosmic-ray event generators, which provide different models 
 of soft hadronic interactions. 

\end{abstract}

\vspace*{2.0cm}

\begin{center}
  Submitted to the European Physical Journal C
\end{center}

\vspace{\fill}

{\footnotesize 
\centerline{\copyright~CERN on behalf of the \lhcb collaboration, license \href{http://creativecommons.org/licenses/by/3.0/}{CC-BY-3.0}.}}
\vspace*{2mm}

\end{titlepage}


\newpage
\setcounter{page}{2}
\mbox{~}
\newpage

\centerline{\large\bf LHCb collaboration}
\begin{flushleft}
\small
R.~Aaij$^{38}$, 
C.~Abellan~Beteta$^{33,n}$, 
A.~Adametz$^{11}$, 
B.~Adeva$^{34}$, 
M.~Adinolfi$^{43}$, 
C.~Adrover$^{6}$, 
A.~Affolder$^{49}$, 
Z.~Ajaltouni$^{5}$, 
J.~Albrecht$^{35}$, 
F.~Alessio$^{35}$
% ...

{\footnotesize \it
% ...
$ ^{5}$Clermont Universit\'{e}, Universit\'{e} Blaise Pascal, CNRS/IN2P3, LPC, Clermont-Ferrand, France\\
$ ^{6}$CPPM, Aix-Marseille Universit\'{e}, CNRS/IN2P3, Marseille, France\\
% ...
$ ^{11}$Physikalisches Institut, Ruprecht-Karls-Universit\"{a}t Heidelberg, Heidelberg, Germany\\
% ...
$ ^{33}$Universitat de Barcelona, Barcelona, Spain\\
$ ^{34}$Universidad de Santiago de Compostela, Santiago de Compostela, Spain\\
$ ^{35}$European Organization for Nuclear Research (CERN), Geneva, Switzerland\\
% ...
$ ^{38}$Nikhef National Institute for Subatomic Physics, Amsterdam, The Netherlands\\
% ...
$ ^{43}$H.H. Wills Physics Laboratory, University of Bristol, Bristol, United Kingdom\\
% ...
$ ^{49}$Oliver Lodge Laboratory, University of Liverpool, Liverpool, United Kingdom\\
% ...
$ ^{n}$LIFAELS, La Salle, Universitat Ramon Llull, Barcelona, Spain\\
% ...
}
\end{flushleft}

\cleardoublepage




\pagestyle{plain} 
\setcounter{page}{1}
\pagenumbering{arabic}



%
\section{Introduction}
\label{sec:intro}
In Quantum Chromodynamics~(QCD), the final state of 
an inelastic hadron-hadron collision can be described by contributions 
from hard and soft scattering occurring between the constituents 
of the hadrons, initial- and final-state (gluon) radiation and the 
fragmentation of the initially coloured partonic final state into 
colour-neutral hadrons. The soft component of a collision is called 
the underlying event. Its precise theoretical description remains
a challenge, while the dynamics of hard scattering processes is 
well described by perturbative QCD. One source of the underlying 
event activity is multi-parton interactions~(MPI). These arise mainly 
in the region of a very low parton momentum fraction,
where parton densities are high so that the
probability of more than a single parton-parton interaction per 
hadron-hadron collision is large. MPI effects become increasingly important 
at LHC collision energies, where inelastic interactions between very soft 
partons are sufficiently energetic to contribute 
to final state particle production~\cite{Bartalini:2011jp}.

MPI phenomena can be probed by measuring in the centre-of-mass system 
the amount of energy created in inelastic hadron-hadron interactions 
at large values of the pseudorapidity $\eta=-\ln[\tan(\theta/2)]$, with $\theta$\/ 
being the polar angle of particles with respect to the beam axis. 
The energy flow is expected to be directly sensitive to 
the amount of parton radiation and MPI~\cite{PhysRevD.36.2019}. 
For a particular pseudorapidity interval with width $\Delta\eta$,
the total energy flow, which is normalised to the number 
of inelastic $pp$\/ interactions $N_{\rm int}$, is defined as
\begin{equation}
\label{eq:EF1}
\frac{1}{N_{\rm int}} \frac{dE_{\rm total}}{d\eta} = \frac{1}{\Delta\eta}\left(\frac{1}{N_{\rm int}}\sum_{i=1}^{N_{\rm part,\eta}}E_{i,\eta}\right)  \;,
\end{equation}
\noindent
where $N_{\rm part,\eta}$ is the total number of stable particles
and $E_{i,\eta}$ is the energy of the individual particles.

In this study, the energy flow is measured in $pp$ collisions at \sqs=~7\tev 
within the pseudorapidity range $1.9<\eta<4.9$. This extends the previous measurements 
that have been made in $p\bar{p}$~\cite{Albajar1990261} and $ep$ collisions~\cite{Adloff:1999ws}
to larger pseudorapidity values and higher centre-of-mass energies, and complements the studies
performed by the CMS and ATLAS collaborations~\cite{Chatrchyan:2011wm,Aad:2012dsa}.
Experimental results are compared to predictions given by 
{\sc Pythia}-based~\cite{Sjostrand:2006za,Sjostrand:2007gs} 
and cosmic-ray event generators~\cite{d'Enterria:2011kw,Ostapchenko:2006aa}, 
which model the underlying event activity in different ways. 
In order to probe various aspects of multi-particle production 
in high-energy hadron-hadron collisions, the measurements are performed 
for the following four classes of events: inclusive minimum-bias, hard scattering, 
diffractive, and non-diffractive enriched interactions.

\section{The \lhcb detector}
\label{sec:lhcb}

The \lhcb detector~\cite{Alves:2008zz} is a single-arm forward
spectrometer with an angular coverage from 10\mrad to 300~(250)\mrad
in the bending~(non-bending) plane, designed for the study of $b$- and $c$-hadrons. 
The detector includes a high precision tracking system consisting 
of a silicon-strip vertex detector~(VELO) surrounding the $pp$\/ interaction region, 
a large-area silicon-strip detector located upstream of a dipole magnet 
with a bending power of about $4{\rm\,Tm}$, and three stations of silicon-strip detectors 
and straw drift tubes placed downstream. The VELO has a larger angular acceptance 
than the rest of the spectrometer, including partial coverage of the backward region. 
It allows reconstruction of charged particle tracks in the pseudorapidity
ranges $1.5<\eta<5.0$ and $-4<\eta<-1.5$. 
The combined tracking system has a momentum resolution $\Delta p/p$ that varies 
from 0.4\% at 5\gevc to 0.6\% at 100\gevc, and an impact parameter resolution 
of 20\mum for tracks with high transverse momentum, \pt.~Charged hadrons are identified 
using two ring-imaging Cherenkov detectors. Photon, electron and hadron
candidates are distinguished by a calorimeter system consisting of
scintillating-pad and preshower detectors, an electromagnetic
calorimeter~(ECAL) and a hadronic calorimeter~(HCAL). The calorimeters
have an energy resolution of $\sigma(E)/E=10\%/\sqrt{E}\oplus 1\%$ and 
$\sigma(E)/E=69\%/\sqrt{E}\oplus 9\%$ (with $E$ in GeV),
respectively.  Muons are identified by a
system composed of alternating layers of iron and multiwire
proportional chambers. 

The trigger consists of a hardware stage, based on information from 
the calorimeter and muon systems, followed by a software stage which 
applies a full event reconstruction. For the minimum-bias data used 
in this analysis, the hardware trigger was accepting all beam-beam crossings, 
while the presence of at least one reconstructed track was required 
in the software stage to record an event. 

\section{Data analysis}
\label{sec:analysis}

\subsection{Data and Monte Carlo samples}
The analysis is performed using a sample of minimum-bias data
collected in $pp$ collisions at \sqs=~7\tev during 
the initial running period of the \lhc with low interaction rate. 
The fraction of bunch crossings with two or more collisions (``pile-up events'')  
is estimated to be approximately $5\%$. The total number of events
available in the sample is $5.8\times10^{6}$, corresponding to 
an integrated luminosity of about 0.1\invnb.

Fully simulated minimum-bias $pp$ events at \sqs=~7\tev were generated
using the \lhcb tune~\cite{LHCb-PROC-2010-056} of the {\sc Pythia}~6.4
event generator~\cite{Sjostrand:2006za}.
Here, decays of hadronic particles are described by \evtgen~\cite{Lange:2001uf}
in which final state QED radiation is generated using \photos~\cite{Golonka:2005pn}.
The interaction of the generated particles with the detector and its
response are implemented using the \geant toolkit~\cite{Allison:2006ve, Agostinelli:2002hh}
as described in Ref.~\cite{Clemencic:LHCbMC}.
Additional Monte Carlo~(MC) samples with fully simulated minimum-bias
$pp$ interactions at \sqs=~7\tev were generated using the Perugia~0 and Perugia~NOCR~\cite{Skands:2009zm}
tunes of {\sc Pythia}~6.4.  These models along with the \lhcb tune use different values for
the MPI energy scaling parameter and MPI \pt cut-off, which entails
a sizeable deviation in the amount of MPI predicted by these tunes.
 
The \lhcb tune utilises the CTEQ6L parton density functions~(PDFs)~\cite{Pumplin:2002vw}, 
while both Perugia tunes use the CTEQ5L PDFs~\cite{Lai:1999wy}. 
Colour reconnection effects are not included in the Perugia~NOCR tune. 
In the MC samples generated with the Perugia~0 and Perugia~NOCR tunes, 
diffractive $pp$\/ interactions are not included, whereas 
the sample generated with the \lhcb tune contains the
contributions from both single and double diffractive processes.
A sample of fully simulated diffractive events generated 
with {\sc Pythia}~8.130~\cite{Sjostrand:2007gs}, which utilises the CTEQ5L PDFs, is used in addition. 
This event generator gives a more accurate description of diffractive $pp$\/
interactions than {\sc Pythia}~6, especially at high-\pt, as it includes the contribution 
from hard diffractive processes, which is absent in {\sc Pythia}~6~\cite{Navin:2010kk}.

In addition to the models above, experimental results are compared to 
generator level predictions given by the {\sc Pythia}~8.135 model with default parameters.
Furthermore, the measurements are compared with predictions given by 
the cosmic-ray interaction models {\sc Epos}~1.99~\cite{Pierog:2009zt}, 
{\sc Qgsjet01}, {\sc QgsjetII-03}~\cite{Ostapchenko:2007qb}, and 
{\sc Sibyll}~2.1~\cite{Ahn:2009wx}, which are widely used
in extensive air shower simulations and are not tuned to LHC data.
These generate inelastic $pp$\/ interactions taking into account 
the contributions from both soft and hard scattering processes. 
Soft contributions are described with Gribov's Reggeon field theory~\cite{Grassberger:1978dw} 
via exchanges between virtual quasi-particle states (Pomerons), while hard processes 
are described by perturbative QCD via exchanges of hard or semi-hard Pomerons.
The predictions given by these models diverge mainly because of different treatments 
of non-linear interaction effects related to parton saturation~\cite{Goncalves:2004ek} 
and shadowing~\cite{PhysRevD.63.096001}.
The {\sc Qgsjet01} model describes hadronic multiple scattering
processes as multiple exchanges of Pomerons without specific treatment of saturation effects.
A distinct feature of the {\sc QgsjetII} model is the treatment of non-linear 
parton effects via Pomeron interactions taking into account all order re-summation 
of the corresponding Reggeon field theory diagrams.  
Based on the dual parton model~\cite{PhysRevD.51.64}, {\sc Sibyll} utilises 
the Lund string model~\cite{Andersson:1983ia} for hadronisation and
describes soft and hard processes using the Pomeron formalism and the
minijet model~\cite{Durand:1987yv}, correspondingly. The treatment of
non-linear effects in this model is based on a simple geometrical approach 
of parton saturation. The {\sc Epos} model takes into account energy-momentum correlations 
between multiple re-scatterings and describes non-linear effects using an
effective treatment of lowest order Pomeron-Pomeron interaction graphs. 
It also accounts for the final state interaction of the produced particles. 

\subsection{Analysis strategy}
The energy flow, as defined in Eq.\,\ref{eq:EF1}, is the energy-weighted 
pseudorapidity distribution of particles, normalised to the number of inelastic interactions 
and the $\eta$-bin size. The measurements are performed in ten equidistant 
pseudorapidity bins of width $\Delta\eta=0.3$ over the range $1.9<\eta<4.9$. 
The primary measurement is the energy flow carried by charged particles (charged energy flow). 
It is performed with reconstructed tracks which contain hits in the VELO 
and downstream tracking stations and have momentum in the range $2<p<1000$\gevc. 
Particle identification is not required in this analysis, as the energy  
is taken from the momentum, neglecting particle masses. 
In order to be able to compare the results of the measurements 
with generator level predictions, the reconstructed charged energy flow 
is corrected for detector effects. The total energy flow is determined 
by using a data-constrained MC estimate of the neutral component 
based on information from the ECAL, while the HCAL is not used. 
Details of the procedure are discussed below. 

\subsection{Event classes}

The event classes studied in this analysis are defined as follows.
Inclusive minimum-bias events are selected by requesting the
presence of at least one track originating from the luminous region 
in order to suppress pollution from beam-gas interactions 
and beam halo related background. 
Events with two or more reconstructed primary vertices are rejected 
to suppress pile-up contamination. To minimise biases on the track multiplicity 
of the event, the information on the primary vertex is not used.
The selected inclusive minimum-bias interactions are further classified as 
hard scattering, diffractive and non-diffractive enriched events
using the following criteria:
\begin{itemize}
\item Hard scattering events: at least one track with $\pt>3$\gevc and $1.9<\eta<4.9$.
\item Diffractive enriched events: no tracks reconstructed with $-3.5<\eta<-1.5$.
\item Non-diffractive enriched events: at least one track reconstructed with $-3.5<\eta<-1.5$.  
\end{itemize}
The selection requirements applied for the last two event classes are motivated by the fact that 
a sizeable rapidity gap is an experimental signature of diffractive processes~\cite{Nurse:2011vt}. 
The level of enrichment of the diffractive and non-diffractive samples was studied in simulation,
by retrieving the {\sc Pythia} process type of the $pp$\/ interaction for every selected 
diffractive and non-diffractive candidate. In the case of the \lhcb tune of {\sc Pythia}~6.4, 
the purities of the selected diffractive and non-diffractive enriched samples 
are found to be about $70\%$\/ and $90\%$\/, respectively. Although the actual percentages 
are only meaningful within the specific model, the study shows that the applied selection criteria 
indeed lead to sizeable enhancement of the respective event classes.
 
To minimise the experimental corrections, the definition of
the event classes at generator level is similar to that at
detector level. Inclusive minimum-bias events at generator level
are selected by requiring the presence of at least one outgoing 
final-state charged particle (lifetime $\tau>10^{-8}$~s) in 
the pseudorapidity range $1.9<\eta<4.9$, but without imposing any
condition on its energy. 
The sample of hard scattering events is selected by requesting at
least one final-state charged particle with \pt$>3$\gevc 
and $1.9<\eta<4.9$. The absence or presence of at least one  
final-state charged particle in $-3.5<\eta<-1.5$\/ is used as criterion 
to select diffractive and non-diffractive enriched events among inclusive
minimum-bias interactions, respectively. For the selected events, 
the energy flow at generator level is determined using 
the outgoing final-state charged and neutral particles\symbolfootnote[1]{These include $\pipm$,~$\Kpm$,~$\epm$,~$\mu^{\pm}$,~$\proton$,~$\antiproton$,~$\g$,~$\neutron$,~$\antineutron$ and $\KL$.} 
which are either prompt, originating directly from the fragmentation, 
or the decay products of unstable particles.
Since neutrinos are not reconstructed by the \lhcb spectrometer 
the energy carried by these particles is not taken into account. 
Only MC events simulated with exactly one inelastic $pp$\/ interaction are considered in this study.

\subsection{Corrections}
\label{subsec:correct}

The reconstructed charged energy flow measured with data is corrected 
for detector effects using bin-by-bin correction factors, 
which are estimated as the ratio of the charged energy flow at generator and 
detector level in simulation for each $\eta$ region and event class under consideration. 
The overall correction factor for each bin is taken as the average of 
the correction factors obtained with different MC models used in this analysis. 
For inclusive, hard scattering and non-diffractive enriched events, 
the average and standard deviation of the correction factors, which is included in the 
model-dependent systematic uncertainty, are determined from the \lhcb, 
Perugia~0 and Perugia~NOCR tunes of {\sc Pythia}~6.4. 
In the case of the diffractive enriched event class, 
only the \lhcb tune and the {\sc Pythia}~8 diffractive simulation are used. 
Except for the lowest $\eta$\/ bin, which suffers from reduced acceptance
for low-\pt particles and thus exhibits large corrections and 
a sizeable model dependence, the correction factors are found to be stable
among the models with a slight rise towards the edges of the
detector acceptance. The majority of the factors are well below two, 
indicating that most of the energy is measured by the detector. 
In the case of diffractive enriched events, the correction factors
obtained with the \lhcb tune are slightly smaller
than unity for some of the bins, i.e. the energy flow at detector level is found 
to be larger than at generator level. This is due to
detection inefficiency for charged particles over
the pseudorapidity range $-3.5<\eta<-1.5$. As a result, some of the events 
containing backward going charged particles migrate into the diffractive sample, 
which leads to enhanced energy flow at detector level.

For the measurement of the total energy flow, the neutral component 
$F_{\rm neut,\eta}$\/ is estimated in the following way. 
To first order $F_{\rm neut,\eta}$ is assumed to be proportional to the corrected charged 
energy flow $F_{\rm char,\eta}$ with a factor $R_{\rm gen,\eta}$\/, 
which is the average ratio of the neutral energy flow to the charged energy flow obtained at generator 
level for each $\eta$\/ bin and event class with different {\sc Pythia} tunes. 
This ratio is found to be rather stable over the entire pseudorapidity range of the measurements 
with only small variations between the {\sc Pythia} tunes. This reflects 
the usage of the same hadronisation mechanism governed in the {\sc Pythia} generator 
by the Lund string model~\cite{Sjostrand:2006za,Sjostrand:2007gs}. 
The latter successfully describes the hadronisation of quarks and gluons emerging 
from high energy interactions and was rigorously tested for high-\pt processes~\cite{Hofmann:1987qk,Akesson:1984iq,Bartel:1981kh}.
The $R_{\rm gen,\eta}$\/ ratio is found to be around $0.6$\/ 
for all event types except the hard scattering interactions. 
For the latter, it is about $15\%$ smaller for all $\eta$ bins. 
This feature is found to be a consequence of the requirement of 
a high-\pt charged particle in the definition of this event class. 

Under the assumption outlined above, the total energy flow for a particular event class 
and pseudorapidity bin $F_{\rm total,\eta}$\/ can be written as
\begin{equation}
\label{eq:EF2}
  F_{\rm total,\eta} = F_{\rm char,\eta}+F_{\rm neut,\eta} = 
  F_{\rm char,\eta} \times \left(1+R_{\rm gen,\eta}\right).  
\end{equation}
In order to constrain this initially purely model-based estimate of
the neutral energy flow to data, the total energy flow is further multiplied 
by an additional correction factor $k_{\eta}$\/. 
It accounts for differences between simulation and data 
being defined for every $\eta$ bin as 
\begin{equation}
\label{eq:EF3}
  k_{\eta} = \frac{1+R_{\rm data,\eta}}{1+R_{\rm mc,\eta}} \;.
\end{equation}
\noindent
Here, $R_{\rm data,\eta}$\/ and $R_{\rm mc,\eta}$\/ are ratios 
of the uncorrected neutral to charged energy flow measured in data and simulation, respectively. 
The $R_{\rm mc,\eta}$ ratio is obtained with different {\sc Pythia} tunes and its average 
is taken for the estimation of the $k_{\eta}$ factors. 
The neutral component of $R_{\rm data,\eta}$\/ and $R_{\rm mc,\eta}$\/ 
is measured using reconstructed photon candidates 
which are selected from neutral clusters in the ECAL with an energy greater than 2\gev and \pt$>0.2$\gevc. 
Since the polar angular coverage of the \ecal begins at about 30\mrad, 
there are no measurements of the neutral energy for the last two $\eta$\/ bins 
($\eta>4.3$). The $k_{\eta}$\/ factors for this pseudorapidity region are 
estimated using a linear extrapolation of the $k_{\eta}$\/ factors obtained 
for the pseudorapidity interval $3.1<\eta<4.3$. 
The bins with $\eta<3.1$\/ are not considered for the extrapolation, 
as these are affected by the detection inefficiency for low-\pt charged particles. 
The latter have a low average momentum in this $\eta$ region and thus are unlikely 
to reach downstream tracking stations.  
Except for the lowest $\eta$ bin, which suffers most from the detection inefficiency 
especially in the case of the diffractive enriched event class, the $k_{\eta}$ factors 
are found to be rather close to unity, reflecting the fact that the ratio of the neutral to
charged energy flow is well simulated at detector level.

\section{Systematic uncertainties}
\label{sec:syst}

The total uncertainties on the results are dominated 
by systematic effects, as the statistical uncertainties are found 
to be negligible for all $\eta$ bins and event classes. The various
contributions to the systematic uncertainties are summarised in
Table~\ref{tab:syst}.

\begin{table}[t!]
\caption{\small Relative systematic uncertainties (in percent) affecting the energy flow measurements 
                for all event classes. The total uncertainties are obtained 
                by adding the individual sources in quadrature.  
                The ranges indicate the variation of the uncertainty as a function of $\eta$.}    
\begin{center}
\begin{tabular}{l|c|c|c|c}
~~~~~~Source of        & Inclusive &   Hard     & Diffractive  & Non-diffractive     \\
~~~~~uncertainty      & minbias   & scattering &  enriched    &   enriched          \\
\hline
Model uncertainty on   & $0.6-9.2$   & $0.7-4.1$  & $16-43$     & $0.7-8.6$    \\
correction factors     &             &            &             &              \\
Selection cuts         & $1.0-4.9$   & $2.7-8.8$  & $0.9-2.8$   & $1.1-5.0$    \\
Tracking efficiency    & $3$         & $3$        & $3$         & $3$          \\
Multiple tracks        & $1$         & $1$        & $1$         & $1$          \\
Spurious tracks        & $0.3-1.2$   & $0.4-1.7$  & $0.2-0.7$   & $0.3-1.2$    \\
Magnet polarity        & ---           & ---      & $2.6-7.7$   & ---          \\ 
Residual pile-up       & $1.7$       & $1.7$      & $1.7$       & $1.7$        \\
\hline
Total on $F_{\rm char,\eta}$   & $3.9-11$  & $4.9-10$     & $16-43$     & $4.0-11$ \\ 
\hline
Variation of $R_{\rm gen,\eta}$  & $0.8-6.1$   & $0.7-2.9$  & $1.5-23$   & $0.9-5.5$  \\ 
and $k_{\eta}$ factors &             &            &             &      \\
Photon efficiency      & $1.4-1.6$   & $1.2-1.3$  & $1.3-2.3$   & $1.3-1.6$  \\ 
ECAL miscalibration    & $<1$        & $<1$       & $<1$        & $<1$  \\ 
\hline
Total on $F_{\rm total,\eta}$ & $4.4-13$  & $5.4-11$   & $17-49$     & $4.4-12$ 
\end{tabular}
\end{center}
\label{tab:syst}
\end{table}

For all event types except hard scattering interactions, the largest uncertainty 
on the charged energy flow arises from the model dependence of the bin-by-bin 
correction factors, which is estimated as the standard deviation of the correction factors 
obtained with different {\sc Pythia} tunes.
Here, the largest impact is at low $\eta$, reaching $9\%$\/ 
for inclusive and non-diffractive enriched events, $4\%$\/ for hard 
scattering interactions and up to $43\%$\/ for diffractive enriched events. 
At large $\eta$ this effect generally drops to about 1--2$\%$\/ for all 
event classes except diffractive enriched interactions for which 
it stays above $15\%$.

Systematic uncertainties related to the track selection requirements 
are estimated by comparing the fraction of the energy flow from tracks which
are rejected by the selection cuts in data and simulation. For the majority of the bins
the resulting systematic uncertainty is found to be less than $4\%$. Only for 
hard scattering events this uncertainty approaches $9\%$ at low $\eta$.

To account for differences between the true tracking efficiency and 
that estimated using simulation, a global $3\%$\/ systematic uncertainty 
is assigned across the entire $\eta$ range following the analysis 
presented in Ref.~\cite{Aaij:2010gn}. 
This applies for all event classes under consideration.

The other tracking related factors having an influence on the charged energy 
flow measurements are contaminations from multiply reconstructed tracks and 
tracks created from random combinations of hits (spurious tracks).
The impact of the former is estimated by removing from the measurement all 
tracks found within the same event with similar momentum vectors. 
It is observed that the charged energy flow drops by less than $1\%$\/ 
for all $\eta$ bins and event classes in case of both data and simulation. For the 
final results, a global $1\%$\/ systematic uncertainty for multiply 
reconstructed tracks is conservatively assigned.~The effect of spurious tracks is estimated in simulation by determining the energy flow 
carried by reconstructed tracks which cannot be associated with particles at generator level
and accounting for the difference between the rate of spurious tracks in data and simulation. 
The corresponding systematic uncertainty is found to vary between $0.2\%$ and $2\%$.

It has been checked that reversing the \lhcb magnet polarity has only an influence  
on the measurements of the charged energy flow for the diffractive enriched event class, 
which mainly consists of low-multiplicity events. 
Here, the corresponding effect is assigned as a systematic uncertainty.  

\begin{table}[t!]
\centering
\caption{\small
   Charged energy flow for all event classes and $\eta$ bins with the corresponding 
   systematic uncertainties. The statistical uncertainties are insignificant and not listed. All values are in GeV per unit pseudorapidity interval.
  }  
\begin{tabular}{c| D{,}{ \pm }{-1} D{,}{ \pm }{-1} D{,}{ \pm }{-1} D{,}{ \pm }{4.4} }
Pseudorapidity
& \multicolumn{1}{c}{Inclusive}
& \multicolumn{1}{c}{Hard}
& \multicolumn{1}{c}{Diffractive}
& \multicolumn{1}{c}{Non-diffractive}\\
range
& \multicolumn{1}{c}{minbias}
& \multicolumn{1}{c}{scattering}
& \multicolumn{1}{c}{enriched}
& \multicolumn{1}{c}{enriched}\\
\hline
$1.9<\eta<2.2$ & 12,1 & 37,4   & 4,2  & 13,1 \\
$2.2<\eta<2.5$ & 16,1 & 50,4   & 5,2  & 17,1 \\
$2.5<\eta<2.8$ & 21,1 & 64,4   & 6,2  & 22,1 \\
$2.8<\eta<3.1$ & 27,1 & 83,5   & 9,3  & 29,1 \\
$3.1<\eta<3.4$ & 35,2 & 105,6  & 12,3 & 38,2 \\
$3.4<\eta<3.7$ & 46,2 & 132,6  & 17,4 & 49,2 \\
$3.7<\eta<4.0$ & 58,2 & 161,8  & 22,5 & 61,2 \\
$4.0<\eta<4.3$ & 73,3 & 194,10 & 31,7 & 77,3 \\
$4.3<\eta<4.6$ & 88,4 & 219,12 & 41,7 & 93,4 \\
$4.6<\eta<4.9$ & 112,5& 256,13 & 57,9 & 118,6 
\end{tabular}
\label{tab:result1}
\end{table}

\begin{table}[h!]
\centering
\caption{\small
  Total energy flow for all event classes and $\eta$ bins with the corresponding 
  systematic uncertainties. The statistical uncertainties are insignificant and not listed. All values are in GeV per unit pseudorapidity interval.
  }  
\begin{tabular}{c| D{,}{ \pm }{-1} D{,}{ \pm }{-1} D{,}{ \pm }{-1} D{,}{ \pm }{4.4} }
Pseudorapidity
& \multicolumn{1}{c}{Inclusive}
& \multicolumn{1}{c}{Hard}
& \multicolumn{1}{c}{Diffractive}
& \multicolumn{1}{c}{Non-diffractive}\\
range
& \multicolumn{1}{c}{minbias}
& \multicolumn{1}{c}{scattering}
& \multicolumn{1}{c}{enriched}
& \multicolumn{1}{c}{enriched}\\
\hline
$1.9<\eta<2.2$&  18,2  &  55,6   &  4,2   &  19,2 \\
$2.2<\eta<2.5$&  26,2  &  77,7   &  6,2   &  28,2 \\
$2.5<\eta<2.8$&  36,2  &  102,7  &  10,3  &  38,2 \\
$2.8<\eta<3.1$&  48,3  &  133,8  &  15,5  &  51,3 \\
$3.1<\eta<3.4$&  60,3  &  164,9  &  20,5  &  64,3 \\
$3.4<\eta<3.7$&  75,3  &  203,11 &  27,6  &  80,4 \\
$3.7<\eta<4.0$&  95,4  &  246,15 &  37,9  &  100,4 \\
$4.0<\eta<4.3$&  118,5 &  296,17 &  50,11 &  125,6 \\
$4.3<\eta<4.6$&  144,7 &  329,20 &  65,11 &  151,7 \\
$4.6<\eta<4.9$&  182,9 &  380,21 &  89,15 &  191,10
\end{tabular}
\label{tab:result2}
\end{table}

Events with more than one reconstructed primary vertex are
vetoed in the analysis in order to suppress pile-up contamination. 
Its residual effect is estimated to be $1.7\%$ by 
taking the efficiencies to accept pile-up events from simulation. This factor 
is included in the normalisation of the energy flow and conservatively 
taken as the systematic uncertainty.

The total energy flow acquires an additional uncertainty 
from the variation of the $R_{\rm gen,\eta}$ and $k_{\eta}$ factors between the {\sc Pythia} tunes 
and the extrapolation procedure used for the $k_{\eta}$ factors in two highest $\eta$ bins. 
No systematic uncertainty is assigned to account for inaccuracies of the Lund string model in describing 
the ratio of the neutral energy flow to the charged energy flow.
The uncertainties associated with the ECAL energy calibration and 
photon reconstruction efficiency also affect the accuracy of the total energy flow.
To account for a possible difference in the photon reconstruction efficiency 
between simulation and data, a global $3.7\%$\/ systematic uncertainty 
is assigned following the analysis presented in Ref.~\cite{LHCb:2012cw}.
The energy calibration of the ECAL has been studied by measuring 
the invariant masses of diphoton resonances~($\pi^{0}\to\gamma\gamma$ and $\eta\to\gamma\gamma$) 
and has been assigned a global systematic uncertainty of $1.5\%$.
Both uncertainties are scaled with a factor 
$F_{\rm neut,\eta}/F_{\rm total,\eta}$ and 
are listed in Table~\ref{tab:syst}.

Other potential sources of systematic uncertainties such as momentum- and 
$\eta$-smearing, effect of the beam crossing angle, neglecting the masses of charged particles, 
pollution from elastic scattering and beam-gas interactions have been studied as well. 
Their impacts on the accuracy of the measurements are found to be negligible.

The total systematic uncertainties on the corrected charged and total 
energy flow are listed in Tables~\ref{tab:result1} and \ref{tab:result2} 
for all event classes and $\eta$ bins. 
It should be noted that the uncertainties 
are strongly correlated between the bins.

\section{Results}
\label{sec:results}

\begin{figure}[htb!]
\centering
\includegraphics[width=0.475\textwidth]{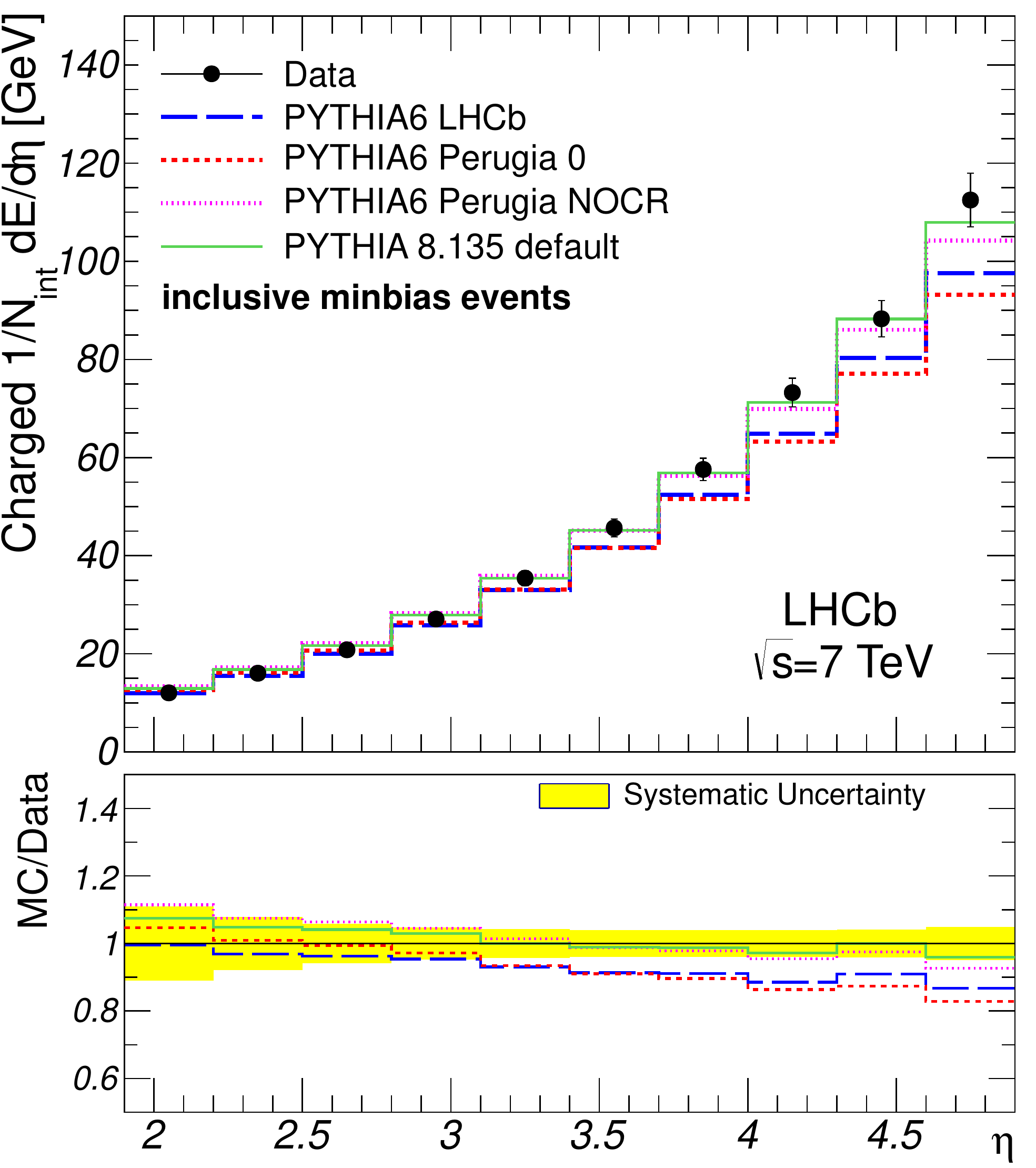}
\includegraphics[width=0.475\textwidth]{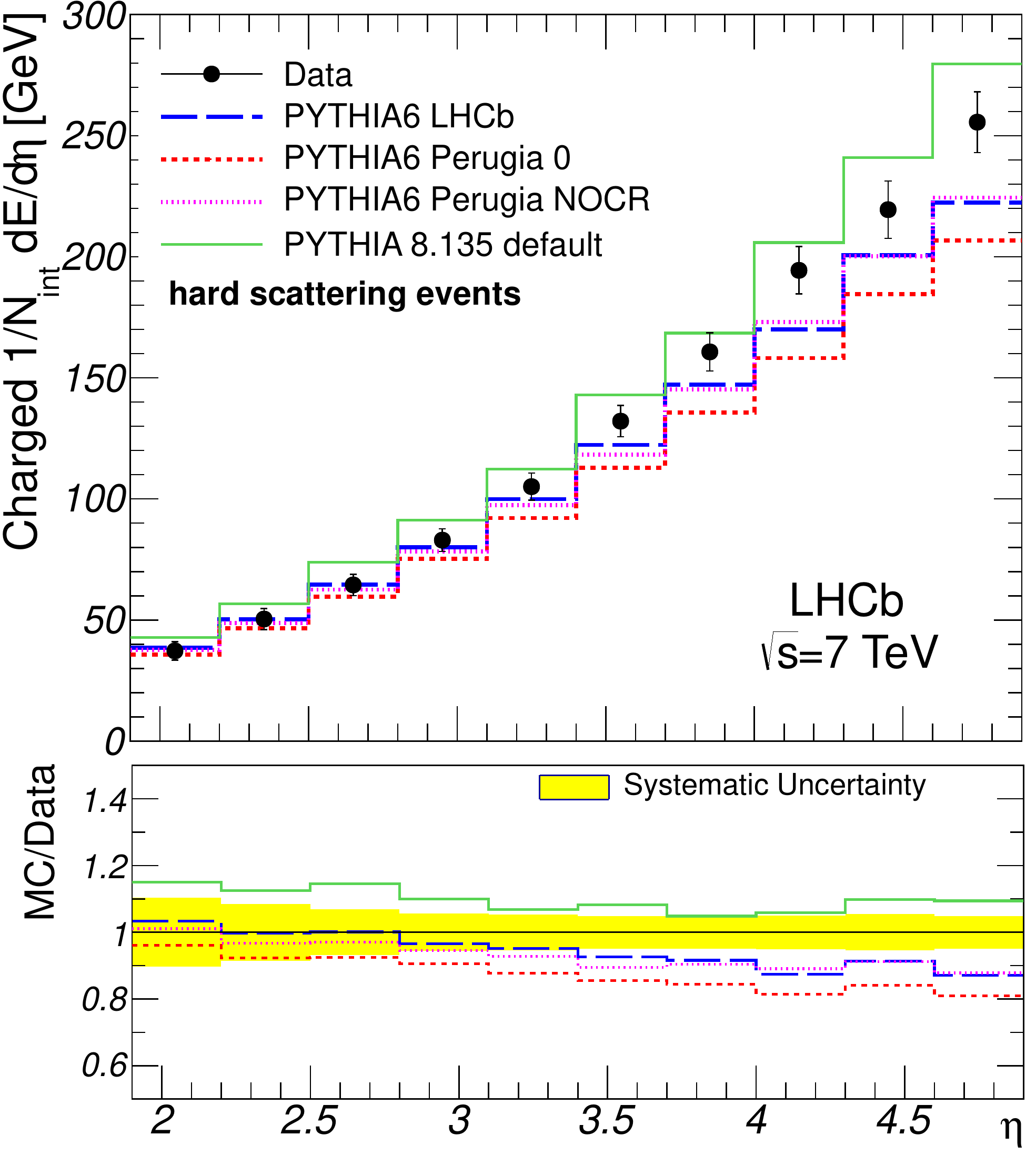}
\includegraphics[width=0.475\textwidth]{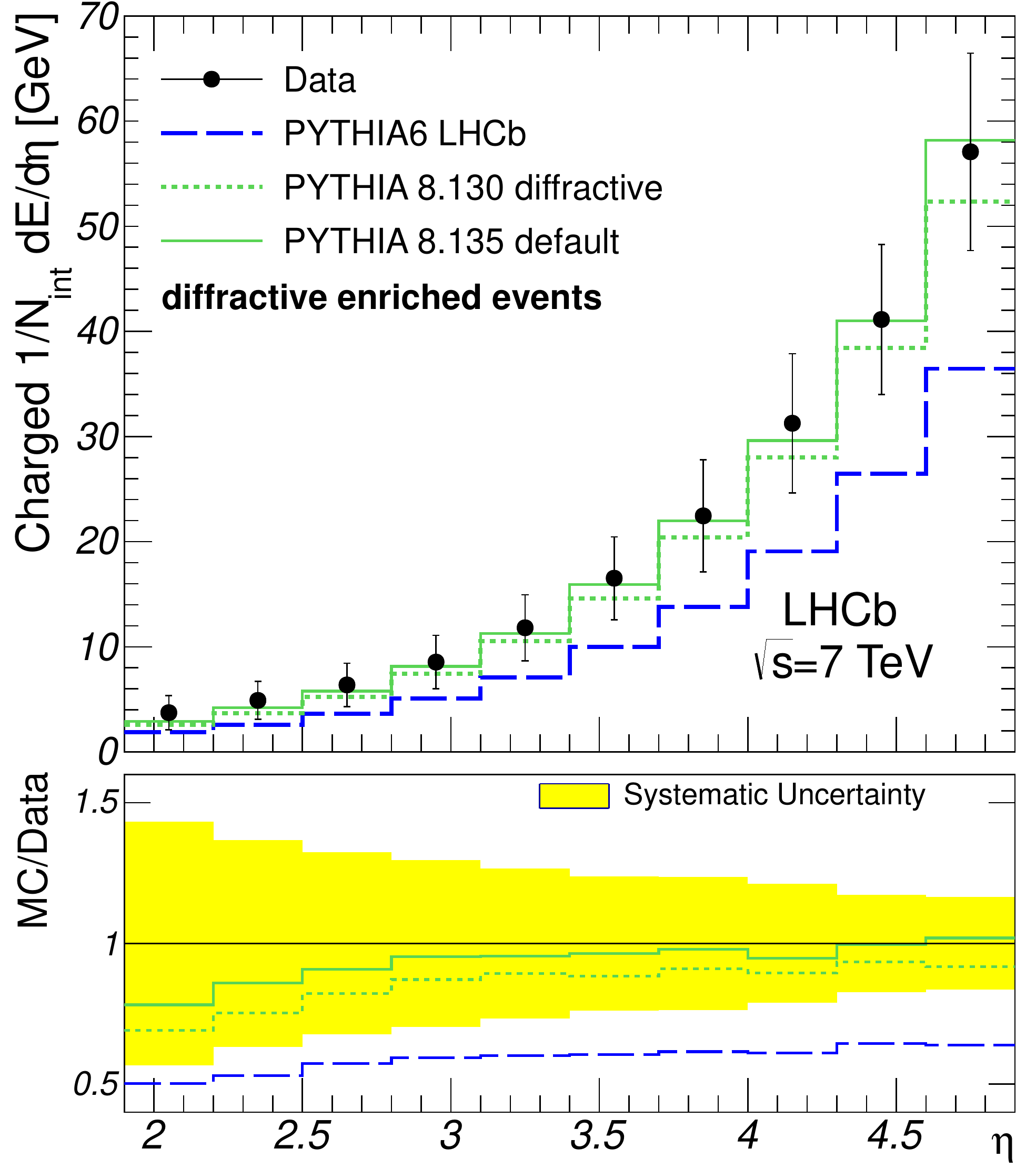}
\includegraphics[width=0.475\textwidth]{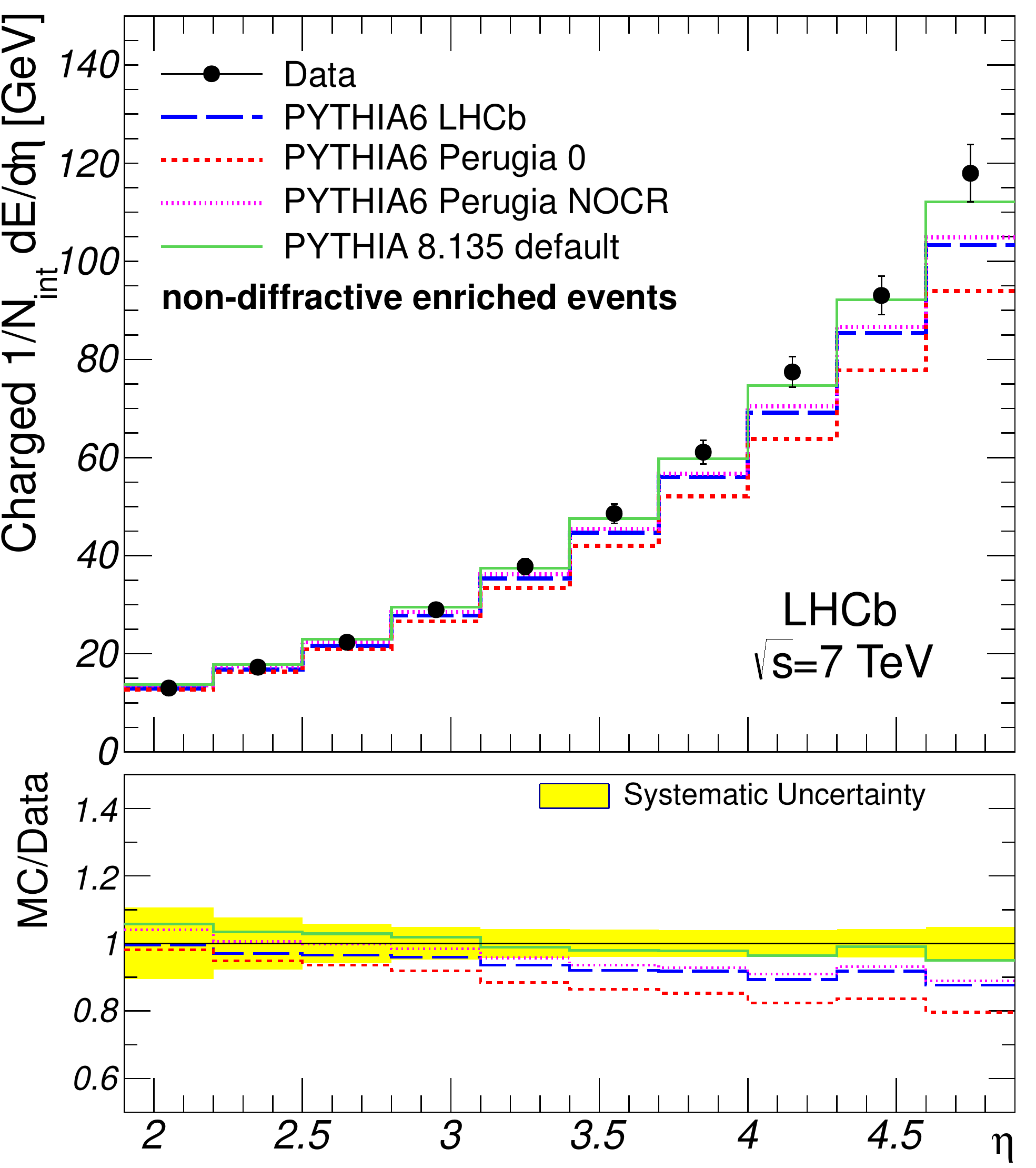}
\vspace*{5mm}
\caption{\small
  Charged energy flow as a function of $\eta$ for all event classes as indicated in the figures. 
  The corrected measurements are given by points with error bars, while the predictions 
  by the {\sc Pythia} tunes are shown as histograms. 
  The error bars represent the systematic uncertainties, which are highly correlated between the bins.
  The statistical uncertainties are negligible. The ratios of MC predictions to data are shown in addition.
}
\label{fig:corrEF} 
\end{figure}

The fully-corrected measurements for the charged energy flow 
are shown in Fig.~\ref{fig:corrEF} for each event class together 
with the generator level predictions given by the {\sc Pythia} tunes
and the corresponding systematic uncertainties.
By comparing experimental results obtained for different event classes
one can clearly see that the amount of energy flow strongly correlates
with the momentum transfer in an underlying $pp$ inelastic interaction.
The charged energy flow rises more steeply with pseudorapidity 
in data than predicted by the majority of the {\sc Pythia} tunes.
As a consequence, the discrepancy between the measurements and
generator level predictions increases towards large $\eta$ rising 
to $20\%$ in the last $\eta$ bin. At lower $\eta$ the data
are reasonably well described by the {\sc Pythia} tunes.
This is the case for all event classes except the diffractive enriched one.
For the latter, 
the measurements are well described by the {\sc Pythia}~8.135 generator with default parameters.
However, this model overestimates the charged energy flow in the case of
hard scattering events over the entire pseudorapidity range of the measurements.

\begin{figure}[htb!]
\centering
\includegraphics[width=0.475\textwidth]{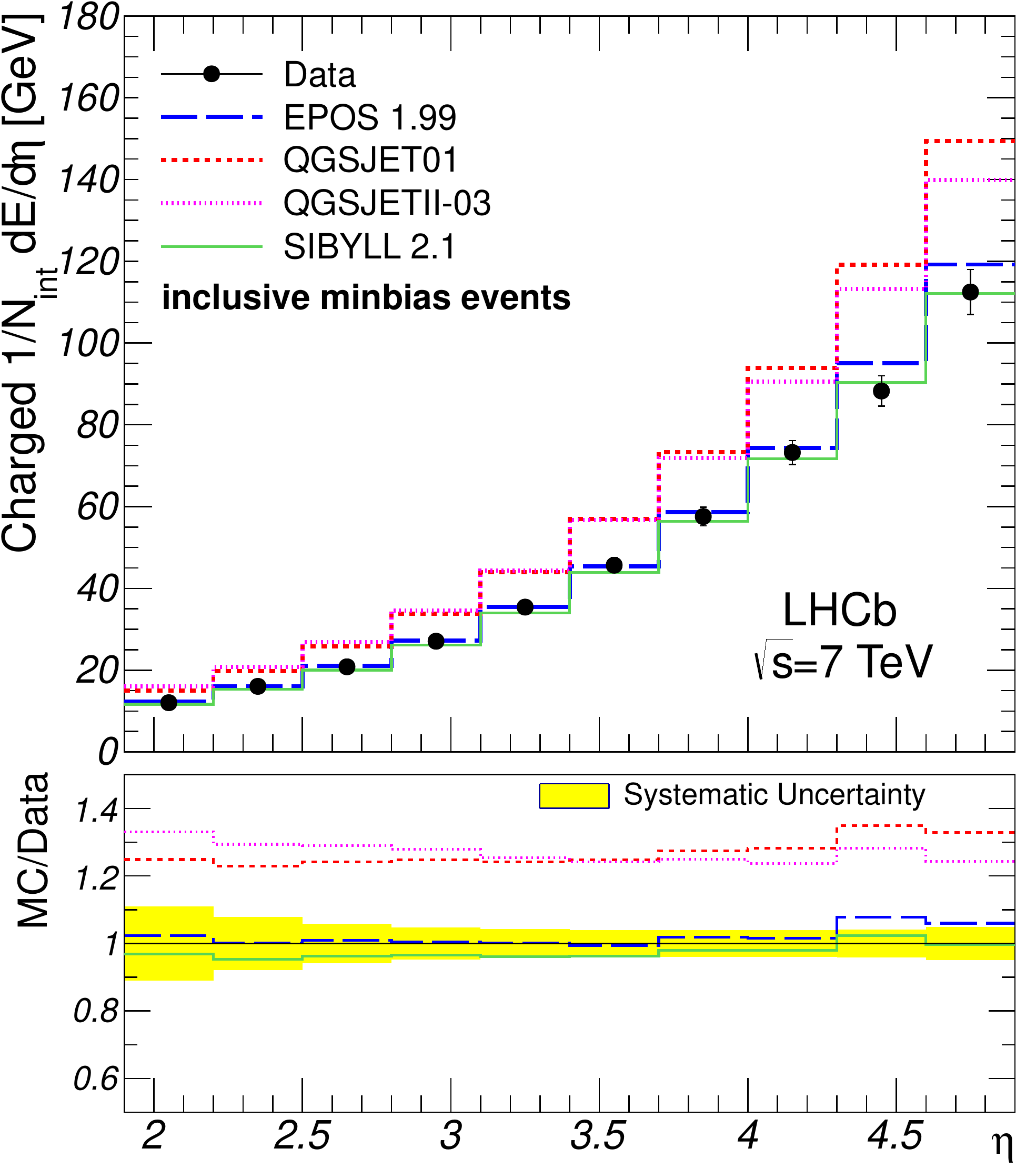}
\includegraphics[width=0.475\textwidth]{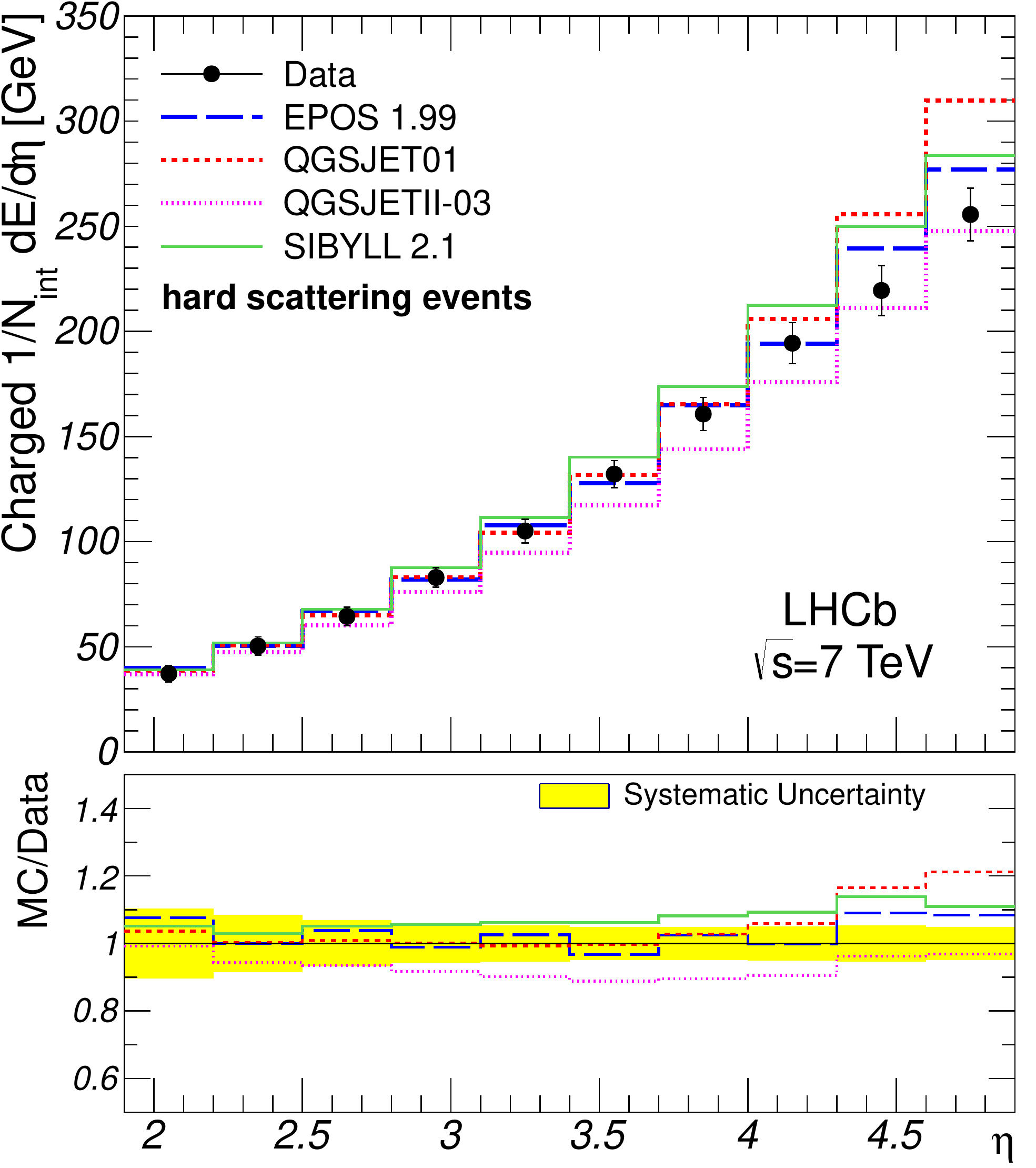}
\includegraphics[width=0.475\textwidth]{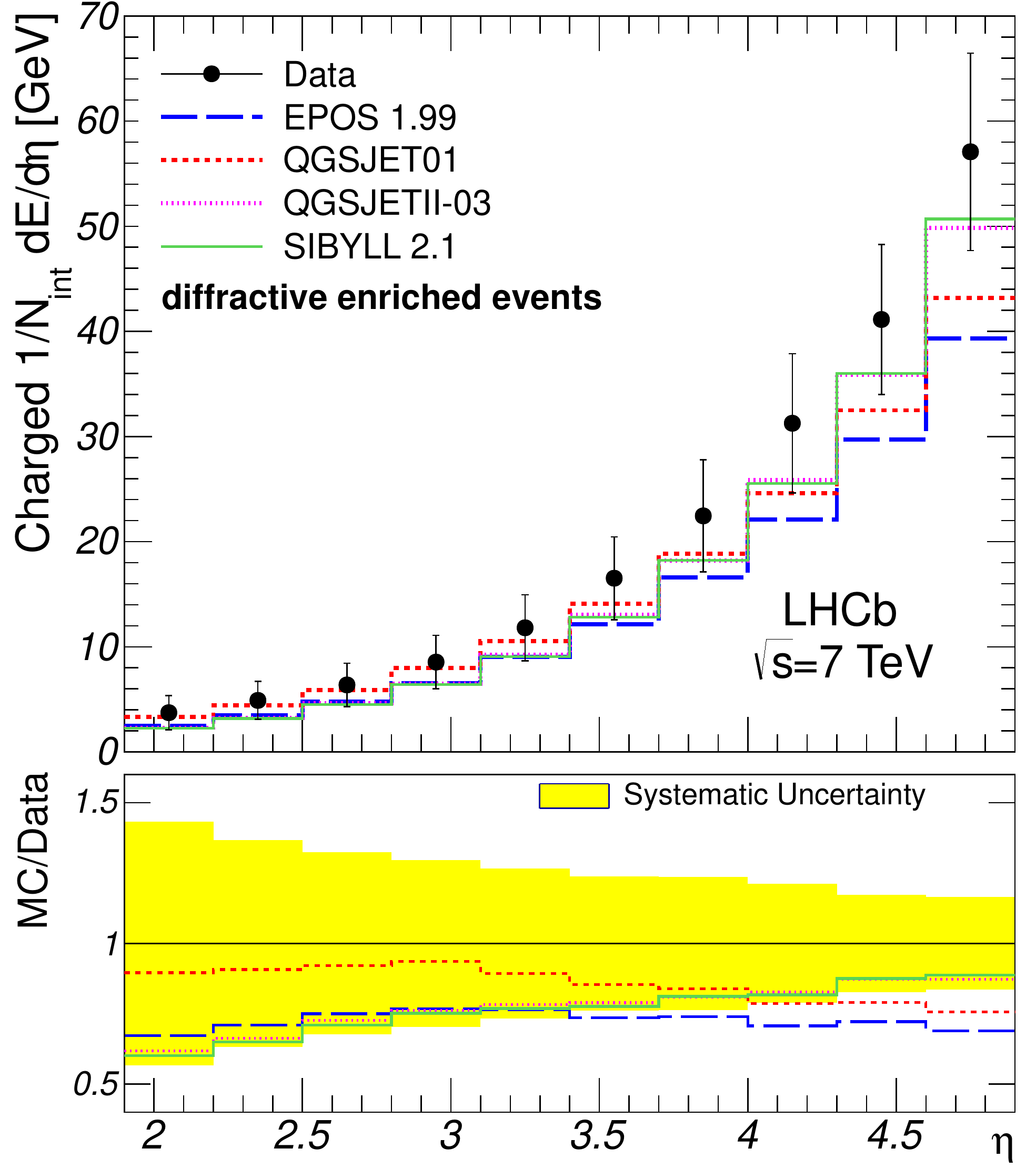}
\includegraphics[width=0.475\textwidth]{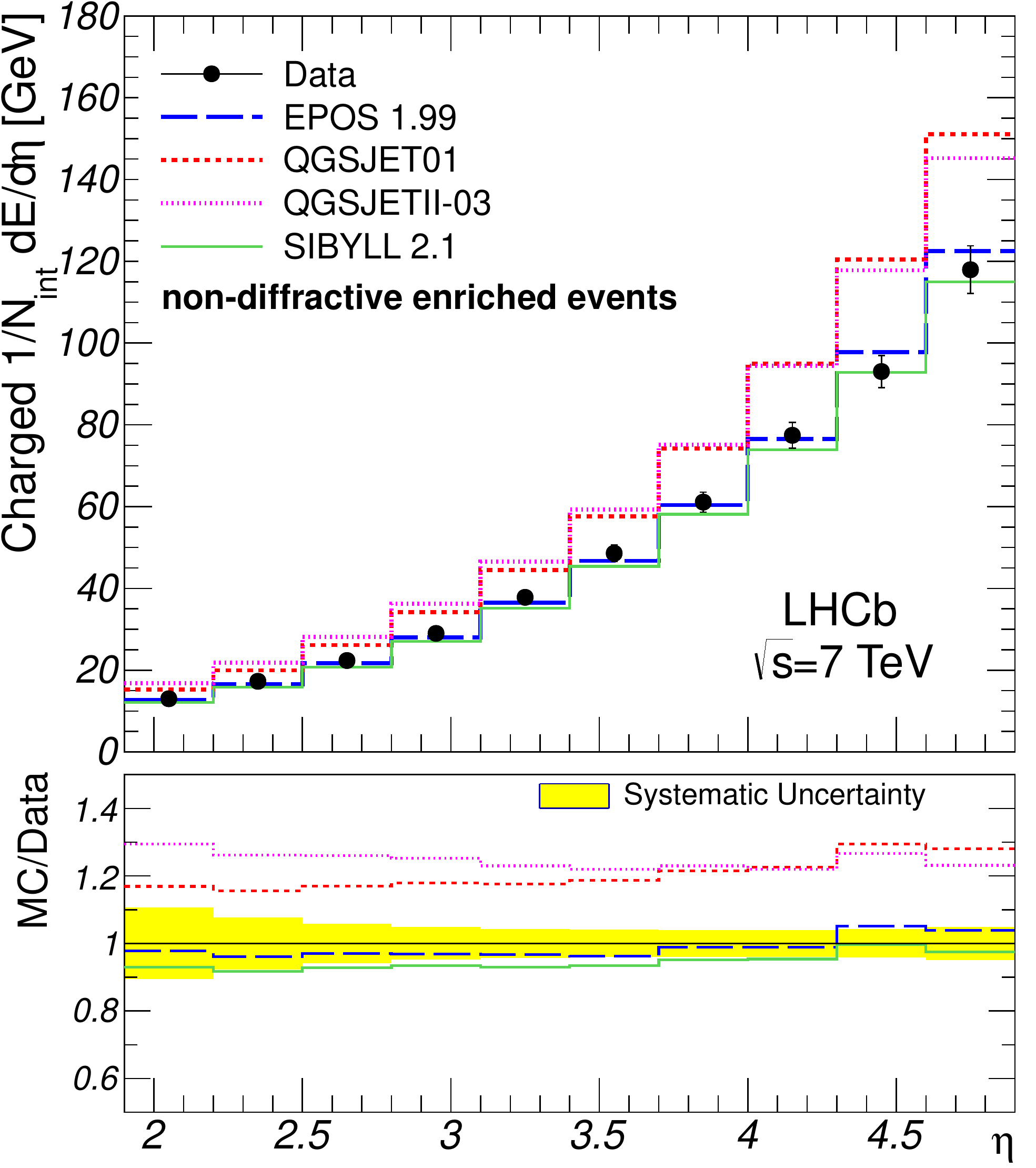}
\vspace*{5mm}
\caption{\small
 Charged energy flow as a function of $\eta$ for all event classes as indicated in the figures.
 The corrected measurements are given by points with error bars, while
 the predictions by the cosmic-ray interaction models are shown as histograms. 
 The error bars represent the systematic uncertainties, which are highly correlated between the bins.
 The statistical uncertainties are negligible. The ratios of MC predictions to data are shown in addition.
}
\label{fig:corrEFc} 
\end{figure}

Figure~\ref{fig:corrEFc} illustrates the charged energy flow
along with the predictions given by the 
cosmic-ray interaction models.
It is interesting to note that the measurements performed with 
inclusive minimum-bias and non-diffractive enriched events are 
well described by the {\sc Epos}~1.99 and  {\sc Sibyll}~2.1 models,
while the {\sc Qgsjet01} and {\sc QgsjetII-03} generators overestimate the charged
energy flow for these event classes. The latter also occurs at large $\eta$ in the case of 
hard scattering interactions for all cosmic-ray interaction models except the {\sc QgsjetII-03}.
The diffractive enriched charged energy flow is underestimated 
by all cosmic-ray interaction models.
\begin{figure}[htb!]
\centering
\includegraphics[width=0.475\textwidth]{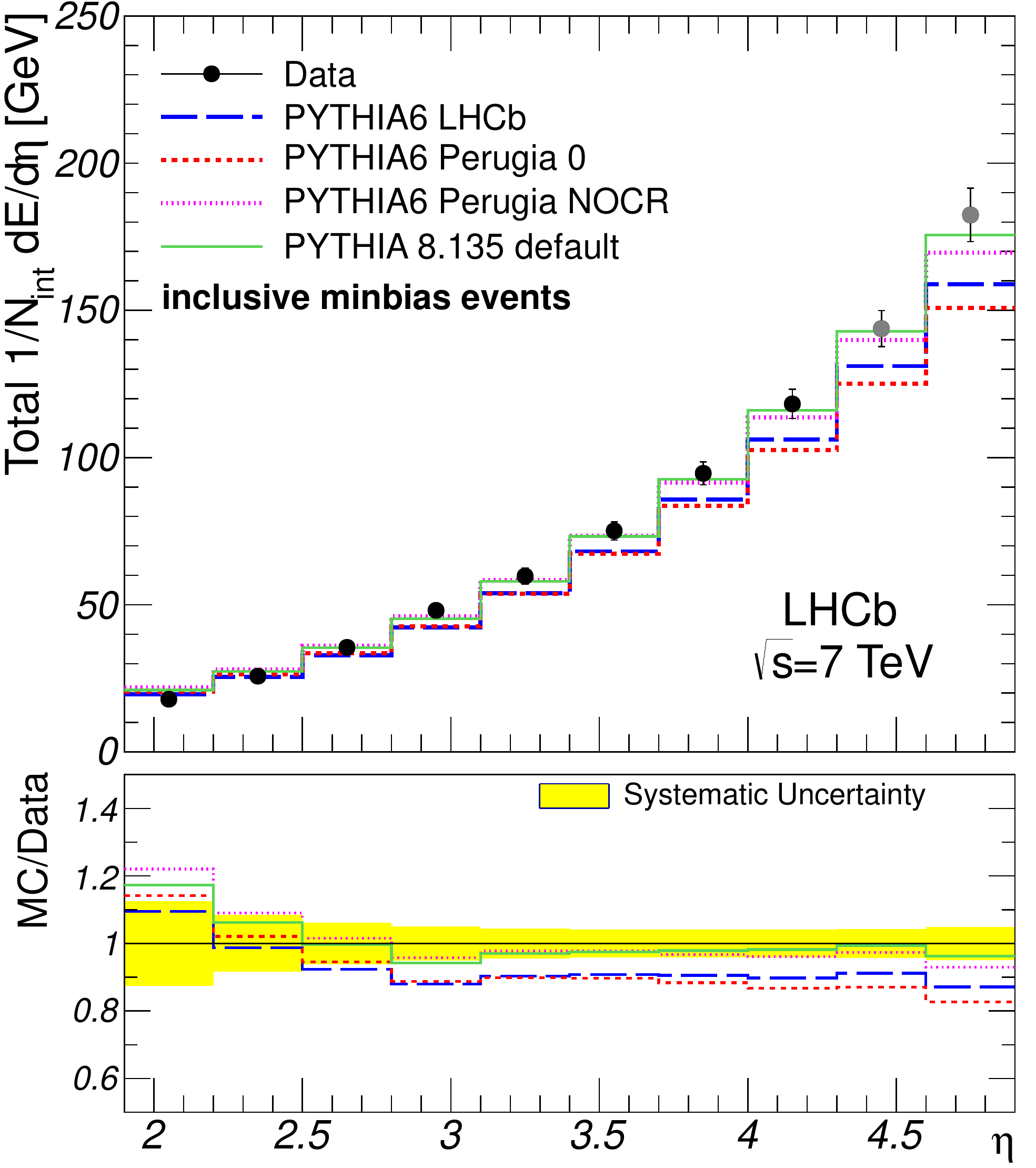}
\includegraphics[width=0.475\textwidth]{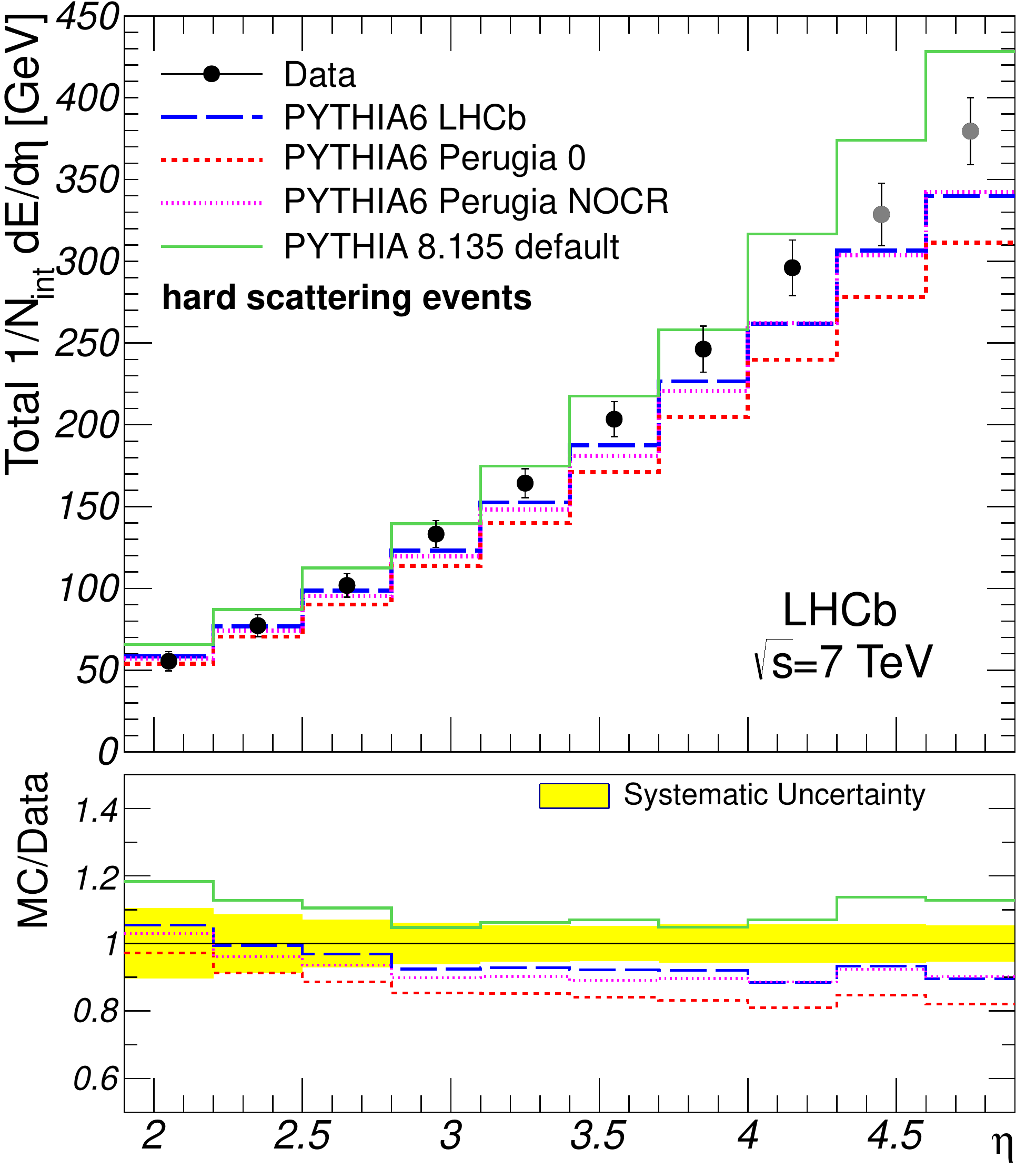}
\includegraphics[width=0.475\textwidth]{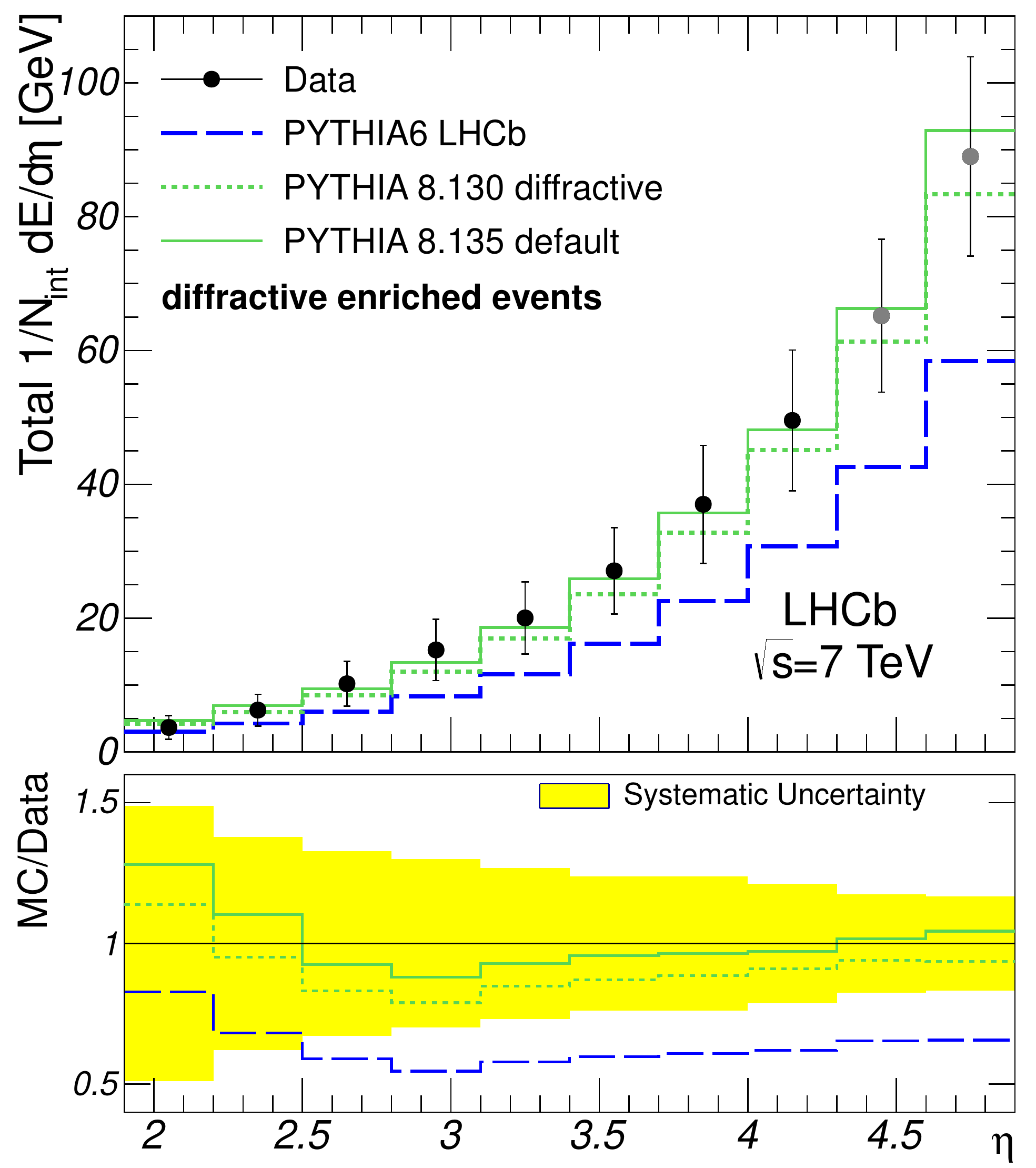}
\includegraphics[width=0.475\textwidth]{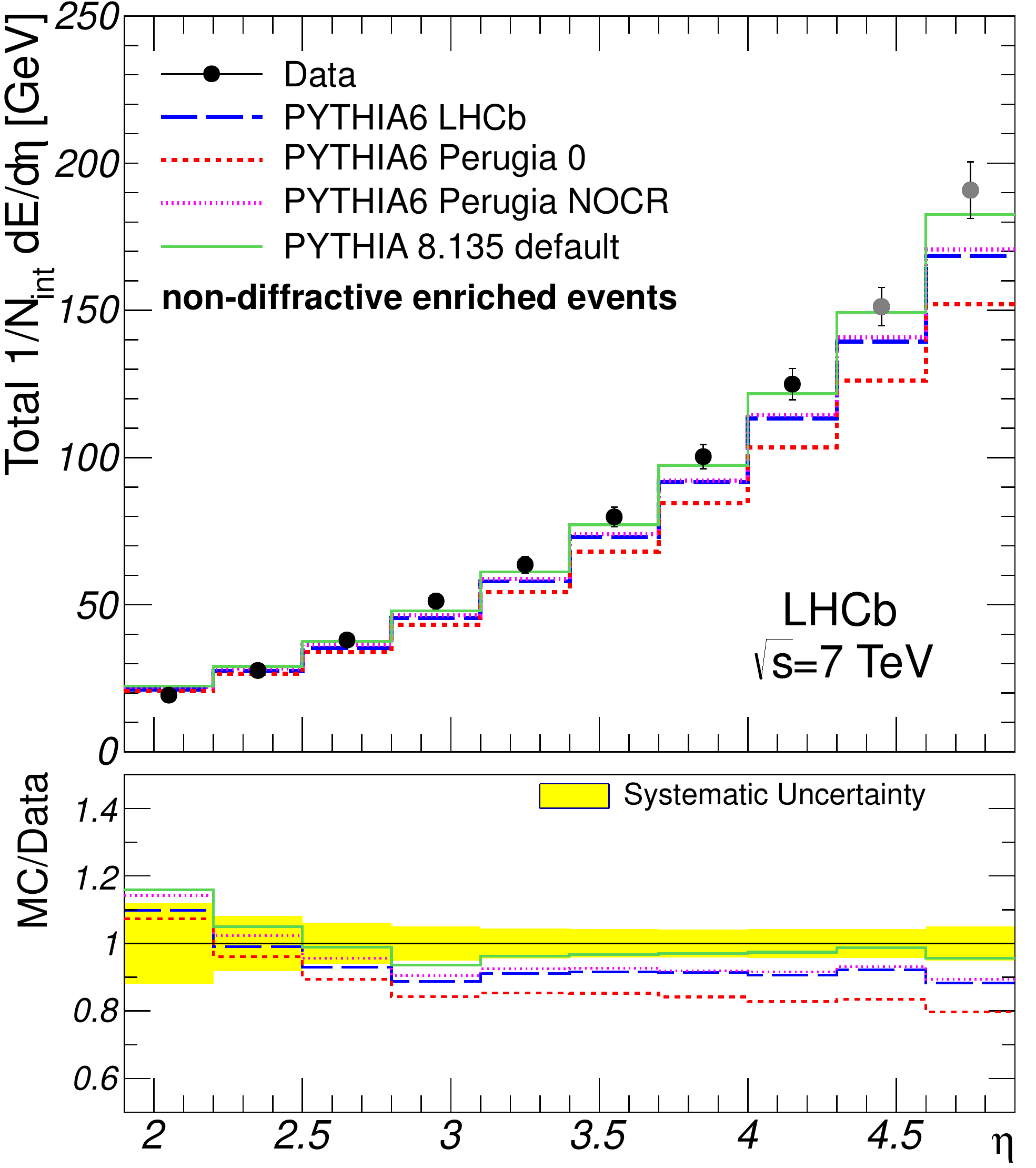}
\vspace*{5mm}
\caption{\small Total energy flow as a function of $\eta$ for all event classes as indicated in the figures. 
             The corrected measurements are given by points with error bars, while the predictions by 
             the {\sc Pythia} tunes are shown as histograms. The data obtained with extrapolated $k_{\eta}$ 
             factors are shown in grey. The error bars represent the systematic uncertainties, which are 
             highly correlated between the bins. The statistical uncertainties are negligible. 
             The ratios of MC predictions to data are shown in addition. 
        }
\label{fig:corrTotal} 
\end{figure}
\begin{figure}[htb!]
\centering
\includegraphics[width=0.475\textwidth]{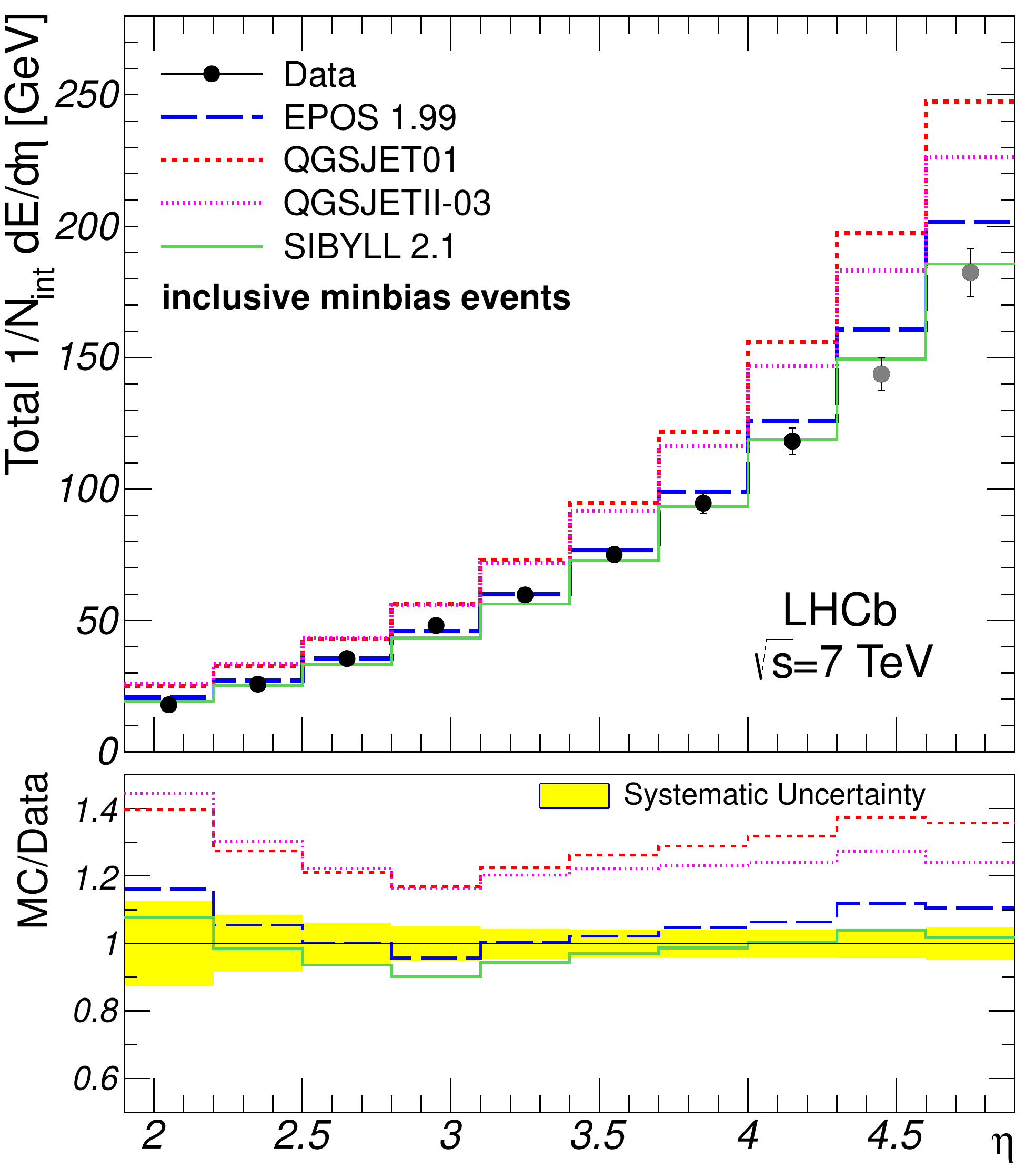}
\includegraphics[width=0.475\textwidth]{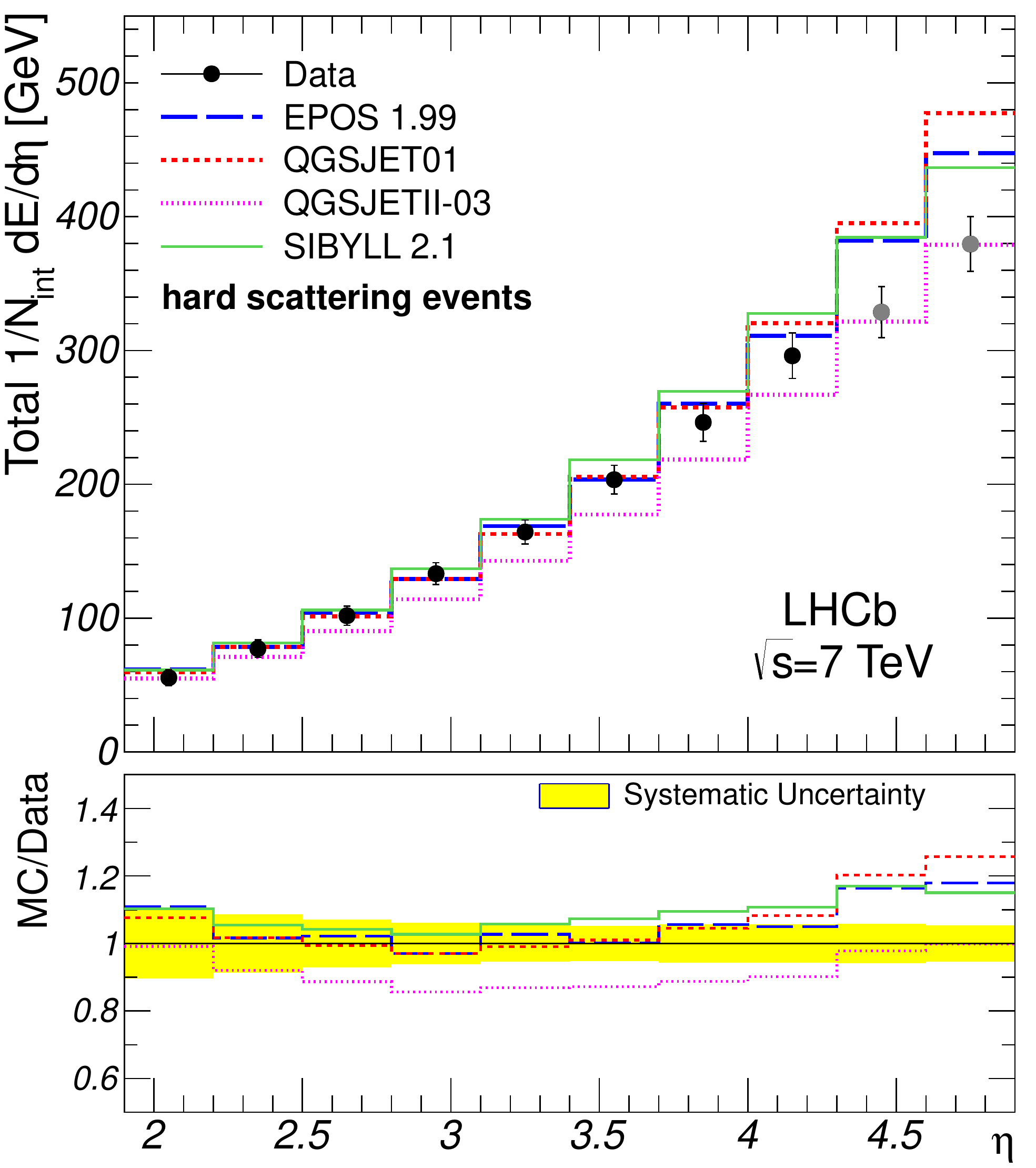}
\includegraphics[width=0.475\textwidth]{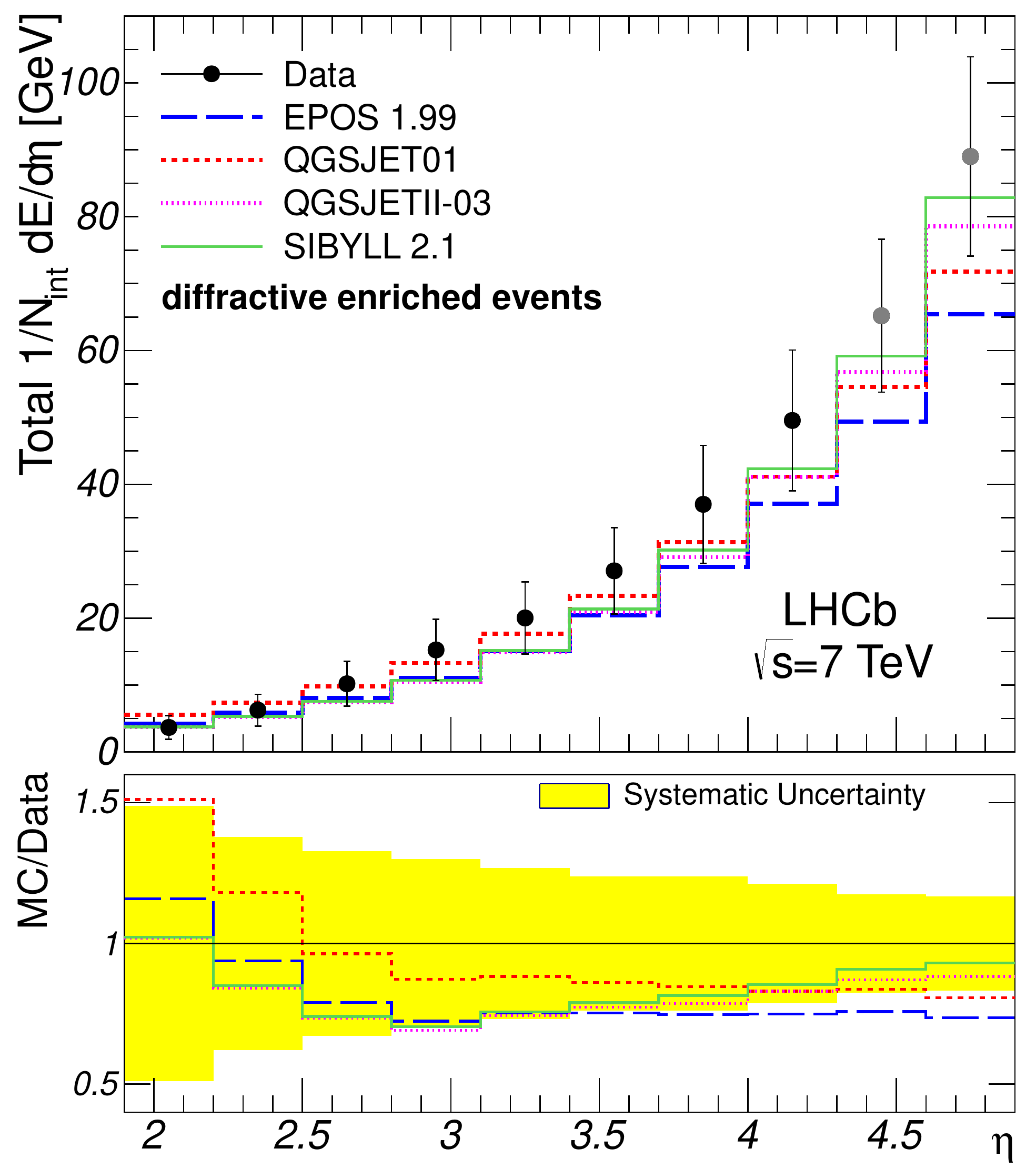}
\includegraphics[width=0.475\textwidth]{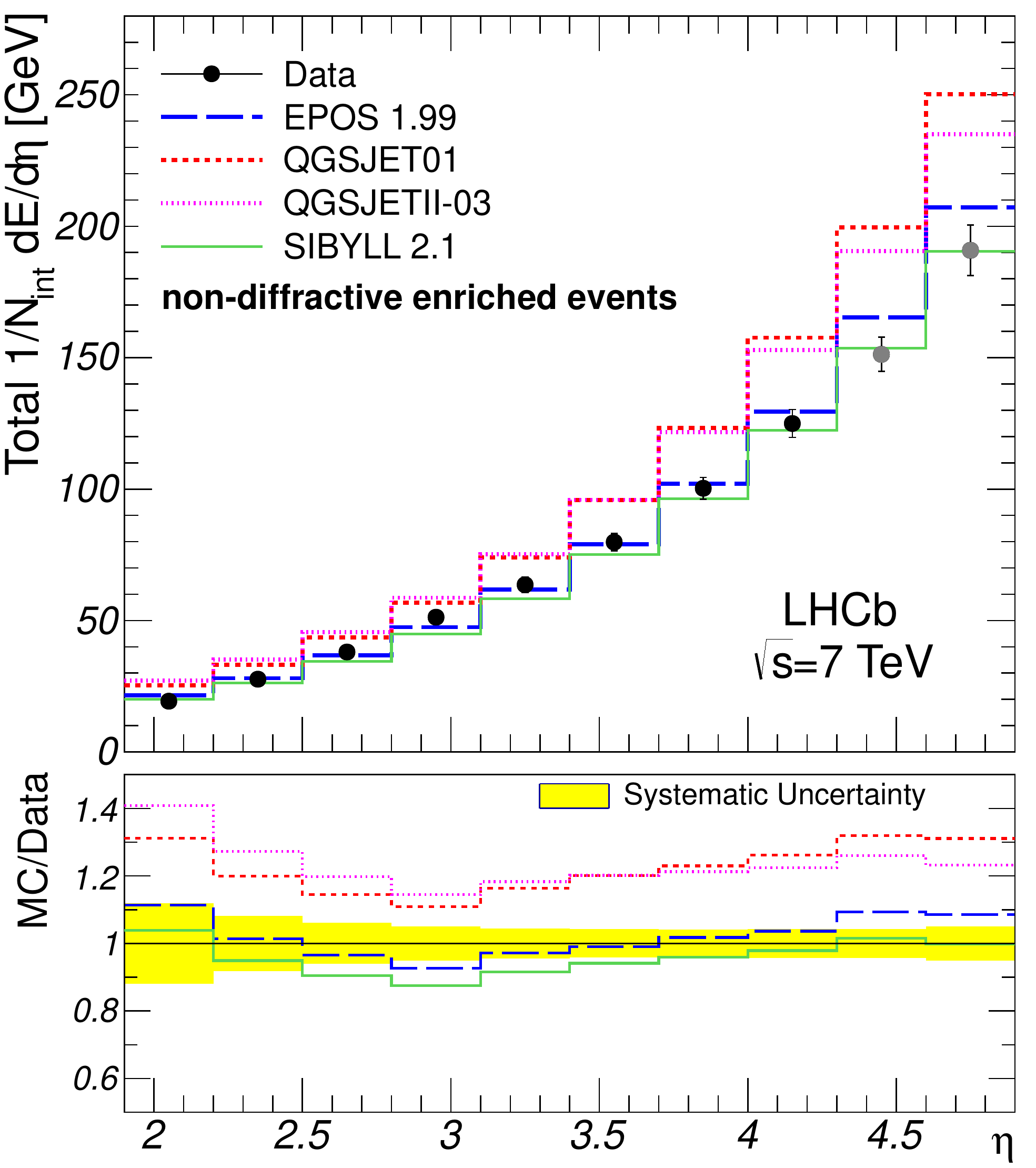}
\vspace*{5mm}
\caption{\small Total energy flow as a function of $\eta$ for all event classes as indicated in the figures. 
             The corrected measurements are given by points with error bars, 
             while the predictions by the cosmic-ray interaction models are shown as histograms.
             The data obtained with extrapolated $k_{\eta}$ factors are shown in grey.
             The error bars represent the systematic uncertainties, which are highly correlated between the bins.
             The statistical uncertainties are negligible. The ratios of MC predictions to data are shown in addition.
	}
\label{fig:corrTotalc} 
\end{figure}

The total energy flow is shown for each event class
in Fig.~\ref{fig:corrTotal} along with the generator level
predictions given by the {\sc Pythia} tunes and the corresponding systematic uncertainties.
It can be clearly seen that all {\sc Pythia}~6 tunes underestimate
the amount of energy flow at large pseudorapidity for all event classes.
The {\sc Pythia}~8.135 generator gives the best description
of the measurements performed with inclusive minimum-bias, 
diffractive and non-diffractive enriched events 
among the {\sc Pythia} tunes, except for the pseudorapidity range $1.9<\eta<2.5$. 
None of these models provide an accurate description of the energy 
flow measured with hard scattering events. The predictions given by the \lhcb and Perugia~NOCR tunes for 
non-diffractive enriched and hard scattering events are rather similar, 
while the Perugia~0 tune significantly underestimates the energy flow 
for all event classes. For diffractive enriched events, 
the inconsistency between the data and the prediction given by the \lhcb tune is found to 
be rather large throughout the entire pseudorapidity range $1.9<\eta<4.9$, 
while the {\sc Pythia}~8.135 generator with default parameters 
gives a good description of the corresponding energy flow at large $\eta$.

Figure~\ref{fig:corrTotalc} illustrates the total energy flow
together with the predictions given by the cosmic-ray interaction models.
It is observed that the {\sc Sibyll}~2.1 generator gives the best description 
of the energy flow measured with inclusive minimum-bias and non-diffractive enriched events 
at large $\eta$. The predictions given by the {\sc Epos}~1.99 generator
for these event classes also describe the measurements reasonably well. 
In the case of hard scattering interactions, the best description of the data 
at large $\eta$ is given by the {\sc QgsjetII-03} generator. The total 
energy flow measured with diffractive enriched events is underestimated 
at large $\eta$ by all cosmic-ray interaction models used in this study. 
The measurements of the charged and total energy flow are summarised 
in Tables~\ref{tab:result1} and \ref{tab:result2} 
for all event classes and $\eta$ bins. 

The results obtained in this study cannot be directly compared with
the measurements performed by the CMS collaboration~\cite{Chatrchyan:2011wm}, 
since different event selection criteria are applied in the analyses.
Nevertheless, both measurements demonstrate that the energy flow is underestimated by
{\sc Pythia}~6 tunes at large pseudorapidity, while the results 
of the ATLAS collaboration indicate that the amount of transverse energy 
is also underestimated by various {\sc Pythia} tunes at large $\eta$~\cite{Aad:2012dsa}. 

\section{Conclusions}
\label{sec:concl}

The energy flow is measured in the pseudorapidity range
$1.9<\eta<4.9$\/ with data collected by the
\lhcb experiment in $pp$\/ collisions at \sqs=~7\tev for
inclusive minimum-bias interactions, hard scattering processes
and events with enhanced or suppressed diffractive contribution.
The primary measurement is the energy flow carried by charged particles.
For the measurement of the total energy flow, a data-constrained MC estimate 
of the neutral component is used. 
The energy flow is found to increase with the momentum transfer
in an underlying $pp$ inelastic interaction.
The evolution of the energy flow as a function of pseudorapidity
is reasonably well reproduced by the MC models.
Nevertheless, the majority of the {\sc Pythia} tunes
underestimate the measurements at large pseudorapidity,
while most of the cosmic-ray interaction models overestimate them,
except for diffractive enriched interactions. 
For inclusive and non-diffractive enriched events, the best description 
of the data at large $\eta$ is given by the {\sc Sibyll}~2.1 
and {\sc Pythia}~8.135 generators. The latter also provides
a good description of the energy flow measured with 
diffractive enriched events, especially at large $\eta$.
The comparison shows that the absence of hard diffractive processes moderates 
the amount of the forward energy flow meaning that their inclusion
is vital for a more precise description of partonic interactions.
It also demonstrates that higher-order QCD effects as contained in the  
Pomeron phenomenology play an important role in the forward region.
None of the event generators used in this analysis are able to 
describe the energy flow measurements for all event classes that 
have been studied.

\newpage


\section*{Acknowledgements}

\noindent 

We are thankful to Colin Baus and Ralf Ulrich from 
the Karlsruhe Institute of Technology for providing 
the predictions of the cosmic-ray Monte Carlo generators. 
We express our gratitude to our colleagues in the CERN
accelerator departments for the excellent performance of the LHC. We
thank the technical and administrative staff at the LHCb
institutes. We acknowledge support from CERN and from the national
agencies: CAPES, CNPq, FAPERJ and FINEP (Brazil); NSFC (China);
CNRS/IN2P3 and Region Auvergne (France); BMBF, DFG, HGF and MPG
(Germany); SFI (Ireland); INFN (Italy); FOM and NWO (The Netherlands);
SCSR (Poland); ANCS/IFA (Romania); MinES, Rosatom, RFBR and NRC
``Kurchatov Institute'' (Russia); MinECo, XuntaGal and GENCAT (Spain);
SNSF and SER (Switzerland); NAS Ukraine (Ukraine); STFC (United
Kingdom); NSF (USA). We also acknowledge the support received from the
ERC under FP7. The Tier1 computing centres are supported by IN2P3
(France), KIT and BMBF (Germany), INFN (Italy), NWO and SURF (The
Netherlands), PIC (Spain), GridPP (United Kingdom). We are thankful
for the computing resources put at our disposal by Yandex LLC
(Russia), as well as to the communities behind the multiple open
source software packages that we depend on.


\nocite{Chatrchyan:2011wm}
\nocite{Albajar1990261}
\nocite{Adloff:1999ws}
\nocite{PhysRevD.36.2019}
\nocite{Bartalini:2011jp}
\nocite{Alves:2008zz}
\nocite{Lai:1999wy}
\nocite{Pumplin:2002vw}
\nocite{Sjostrand:2006za}
\nocite{Sjostrand:2007gs}
\nocite{Navin:2010kk}
\nocite{Nurse:2011vt}
\nocite{Pierog:2009zt}
\nocite{Ahn:2009wx}
\nocite{Ostapchenko:2007qb}
\nocite{d'Enterria:2011kw}
\nocite{Ostapchenko:2006aa}
\nocite{Durand:1987yv}
\nocite{PhysRevD.51.64}
\nocite{Grassberger:1978dw}
\nocite{Aaij:2010gn}
\nocite{Agostinelli:2002hh}
\nocite{Lange:2001uf}
\nocite{Skands:2009zm}
\nocite{Goncalves:2004ek}
\nocite{PhysRevD.63.096001}
\nocite{Golonka:2005pn}
\nocite{LHCb-PROC-2010-056}
\nocite{Allison:2006ve}
\nocite{Andersson:1983ia}
\nocite{Bartel:1981kh}
\nocite{Hofmann:1987qk}
\nocite{Akesson:1984iq}
\nocite{LHCb:2012cw}
\nocite{Aad:2012dsa}
\bibliographystyle{LHCb}
\bibliography{main}

\end{document}